\documentclass[twocolumn,prd,superscriptaddress,preprintnumbers,amsmath,amssymb,nofootinbib]{revtex4}
\newcommand{\bg}{\bar{g}}
\usepackage{amsmath}
\usepackage{amssymb}
\usepackage{slashed}
\usepackage{graphicx}
\graphicspath{{images/}}
\usepackage{graphics}
\usepackage{epsfig}
\usepackage{color}
\usepackage{tabularx}
\usepackage[section,subsection,subsubsection]{extraplaceins}

\usepackage{hyperref}
\newcommand{\be}{\begin{equation}}
\newcommand{\ee}{\end{equation}}
\newcommand{\bea}{\begin{eqnarray}}
\newcommand{\eea}{\end{eqnarray}}
\newcommand{\Eqref}[1]{Eq.~\eqref{#1}}
\newcommand{\red}[1]{#1}

\begin{document}
\title{Matter matters in asymptotically safe quantum gravity}

\author{Pietro Don\`a}
\email[]{pietro.dona@sissa.it}
\affiliation{International School for Advanced Studies, via Bonomea 265, 34136 Trieste, Italy\\
and INFN, Sezione di Trieste}

\author{Astrid Eichhorn}
\email[]{aeichhorn@perimeterinstitute.ca} 
\affiliation{Perimeter Institute for Theoretical Physics, 31 Caroline Street N, Waterloo, N2L 2Y5, Ontario, Canada}

\author{Roberto Percacci}
\email[]{percacci@sissa.it} 
\affiliation{International School for Advanced Studies, via Bonomea 265, 34136 Trieste, Italy\\
and INFN, Sezione di Trieste}

\begin{abstract} 
We investigate the compatibility of minimally coupled scalar, fermion and gauge fields with asymptotically safe quantum gravity, using nonperturbative functional Renormalization Group methods.
We study $d=4, 5$ and 6 dimensions and within certain approximations find that for a given number of gauge fields there is a maximal number of scalar and fermion degrees of freedom compatible with an interacting fixed point at positive Newton coupling. 
The bounds impose severe constraints on grand unification with fundamental Higgs scalars.
Supersymmetry and universal extra dimensions are also generally disfavored.
The standard model and its extensions accommodating right-handed neutrinos, the axion and dark-matter models with a single scalar are compatible with a fixed point.

\end{abstract}

\maketitle

\section{Introduction}

A quantum theory of gravity that is a viable description of the microscopic dynamics 
of our universe must include dynamical matter degrees of freedom.
Nevertheless, matter is often ignored in quantum gravity, 
or not included in a fully dynamical way. While this could result in a self-consistent
quantum theory of gravity, it is not clear whether it can yield a model of quantum gravity applicable to our universe.
As in other settings, 
where the addition of further degrees of freedom can fundamentally alter 
the character of the theory, 
dynamical matter might not be easily incorporated into a consistent microscopic
description of gravity. 
As an example, consider Yang-Mills theory, which is ultraviolet (UV) complete due to 
asymptotic freedom. 
If too many fermions are present, asymptotic freedom is destroyed.
A similar effect could occur in gravity, where \red{too many} matter degrees of freedom could preclude a particular scenario for a UV completion. 
Here, we will make a first step towards a consistent quantum description of 
matter and gravity, by studying the asymptotic safety scenario for gravity and matter.

The asymptotic safety scenario, proposed in \cite{Weinberg:1980gg} (for reviews see \cite{Nagy:2012ef,Reuter:2012xf,Reuter:2012id,Percacci:2011fr,Litim:2011cp,Litim:2008tt,Percacci:2007sz,Niedermaier:2006ns,Niedermaier:2006wt}) provides an ultraviolet (UV) completion for gravity in the form of an interacting fixed point of the Renormalization Group (RG) flow. 
Let $g_i(k)$ be the couplings, depending on some momentum scale $k$,
$\beta_{g_i}= k \, \partial_k g_i(k)$ the corresponding beta functions,
 and
\be
\tilde g_i(k) = g_i(k) k^{-d_i}
\ee
the dimensionless form of the couplings ($d_i$ being the mass dimension of $g_i$).
The $\beta$ functions for the dimensionless couplings  $\tilde g_i$
depend on the $\tilde g_n$ themselves, but not explicitly on the RG scale $k$:
\be
\beta_{\tilde g_i}(\tilde g_n)
= \partial_t \tilde g_i(k):=k \partial_k \tilde g_i(k)= -d_i \tilde g_i(k) + f(\tilde g_n).
\ee
The first term shows the scale-dependence due to the canonical dimensionality, which would be present also in a classical theory, whereas the second term arises from quantum fluctuations that introduce a non-trivial scale dependence.
A fixed point is a simultaneous zero of the beta functions of all the dimensionless
couplings: $\beta_{\tilde g_i}(\tilde g_{j*})=0$.
For a Gau\ss{}ian fixed point, where the theory becomes asymptotically free, 
$\tilde g_{i\, \ast}=0$.
This fixed point is known not to lead to a consistent UV limit for gravity, as the Newton coupling by itself cannot become asymptotically free due to its canonical dimensionality, and the addition of curvature-squared couplings leads to an asymptotically free but non-unitary theory \cite{Stelle:1976gc, Stelle:1977ry}.
Thus a quantum field theoretic UV completion of gravity demands $g_{i\, \ast}\neq0$.

An important technical advantage of this scenario lies in its formulation as a local 
continuum quantum field theory. Many powerful tools, which have been successfully
used to describe the other interactions, are available in this setting.
In particular the inclusion of matter is in  principle straightforward, 
and there is no difficulty in considering, e.g., chiral fermions 
\cite{Vacca:2010mj, Eichhorn:2011pc,Gies:2013noa}, 
in contrast to several other approaches to quantum gravity. 

The main advantage of asymptotic safety, however, is predictivity.
Although infinitely many couplings are expected to be nonvanishing at the fixed point, 
predictivity can be retained in this scenario: This happens if all but a finite number of
couplings are UV-repulsive. These irrelevant couplings need to be fine-tuned so that the IR-starting point for the RG flow (when followed ``backwards'' towards the UV) lies within the UV critical surface. This amounts to a prediction of the value of the irrelevant couplings and will yield a predictive theory if the critical surface is finite dimensional.

To determine whether a coupling is relevant, 
we linearize the RG flow  of the dimensionless couplings around the fixed point up to first order
\be
\beta_{\tilde g_i}(\{\tilde g_n\})= 0 + \sum_j\frac{\partial \beta_{\tilde g_i}}{\partial \tilde g_j} \Big|_{\tilde g_n = \tilde g_{n\, \ast}}\left(\tilde g_j - \tilde g_{j\, \ast} \right) +...\,.\label{linflow}
\ee
The solution to this equation can easily be found, and is given by
\be
\tilde g_{i}(k) = \tilde g_{i\, \ast} + \sum_I C_I V^{I}_i  \left(\frac{k}{k_0}\right)^{- \theta_I},
\ee
where 
\be
- \frac{\partial \beta_{\tilde g_i}}{\partial \tilde g_j} \Big|_{g_n = g_{n\, \ast}} V_I = \theta_I V_I.
\ee
Here, the $C_I$'s are constants of integration, which constitute the free parameters of the theory, and $k_0$ is an arbitrary reference momentum scale. The $V_I$ are the eigenvectors, in general given by a linear superposition of couplings, and the critical exponents $-\theta_I$ the eigenvalues of the stability matrix, defined in \eqref{linflow}. At a Gau\ss{}ian fixed point, the critical exponents are given by the mass dimensions of the couplings. Clearly the requirement to hit the fixed point in the UV translates into the condition $C_I=0$ for those $I$ such that $Re[\theta_I]<0$, i.e., the UV-repulsive directions. In contrast, the $C_I$ are undetermined if $Re[\theta_I]>0$, as those directions approach the fixed point automatically towards the UV.  To fix these constants, the values of couplings have to be measured in an experiment, just as is the case of power-counting relevant operators in an asymptotically free theory.
\red{For} instance the gauge coupling in QCD is a free parameter of the standard model. 
Therefore a predictive theory can only have a finite number of positive critical exponents.

The concept of nonperturbative renormalizability, a.k.a. asymptotic safety, is well-known from condensed-matter physics, where the Wilson-Fisher fixed point and its generalizations are interacting fixed points, and their critical \red{exponents} have been shown experimentally to govern second-order phase transitions in a variety of materials. 
In high-energy physics, the concept is applicable to perturbatively nonrenormalizable effective field theories, where it could provide predictive UV completions, see, e.g., \cite{Gies:2009hq,Gies:2013pma}.
Evidence for the existence of a gravitational fixed point with a finite number
of relevant directions has been collected with a variety of tools,
but in recent years most progress has come from functional Renormalization Group methods,
whose application to gravity has been pioneered in \cite{Reuter:1996cp}. 
Various nonperturbative truncations of the RG flow indicate the existence 
of an interacting fixed point 
\cite{Reuter:2001ag,Lauscher:2002sq,
Litim:2003vp,
Fischer:2006fz,
Codello:2006in,
Machado:2007ea,
Codello:2008vh,
Eichhorn:2009ah,
Benedetti:2009rx,
Manrique:2011jc,
Rechenberger:2012dt,
Benedetti:2012dx,
Dietz:2012ic,
Codello:2013fpa, 
Falls:2013bv,
Ohta:2013uca}. 
Recent results also suggest the existence of an infrared fixed point \cite{Donkin:2012ud,Nagy:2012rn, Christiansen:2012rx}.

For a realistic implementation of the asymptotic safety scenario a fixed point for 
pure gravity is not enough:
all matter couplings must also simultaneously reach a fixed point.
Such a study may seem to be hopelessly complicated, but there are
several arguments for pursuing it even at this early stage.
One is the general philosophy, which is widely held in the particle physics
community, that a consistent theory of gravity requires the inclusion of matter.
We will see that the available evidence for asymptotic safety is not particularly
supportive of this point of view, in the sense that a fixed point seems to exist
also for pure gravity, but it is still too early to tell.
Then, even assuming that a pure gravity fixed point existed, 
its applicability to the real world 
would require that matter becomes noninteracting in the ultraviolet, 
and quantum fluctuations of matter do not change the microscopic dynamics of gravity. 
Within a path-integral approach to quantum gravity, it seems more likely that 
quantum fluctuations of all fields determine the microscopic dynamics and drive 
the Renormalization Group flow. This view is supported by the result that 
quantum gravity fluctuations, parametrized as metric or vielbein fluctuations 
in a continuum setting, generate matter interactions even in a free theory 
\cite{Vacca:2010mj, Eichhorn:2011pc, Eichhorn:2012va,Eichhorn:2013ug}.

Other arguments are of a negative character. It is not likely that a firm
proof of (non-)existence  of a gravitational fixed point can be reached.
By widening the scope of the exploration one also enhances the chances of disproving this
scenario. For example, if standard model matter turned out to be incompatible
with a fixed point within the presently available approximations, the case for
asymptotic safety would be correspondingly weakened.
On the experimental side, barring possible surprises, it does not seem very likely that we will
see signatures of Planck scale physics within the foreseeable future.
On the other hand, much more data is expected to become available in particle physics,
and it is possible that some signs in favor of, or against, asymptotic safety
can be found in them.
A striking example is the relatively successful prediction of the Higgs mass
by Shaposhnikov and Wetterich \cite{Shaposhnikov:2009pv}.
Putting that prediction on a firm theoretical basis will require much work on the
mutual influence of gravity and matter at high energy. 

Here we will investigate the consistency of the interacting 
fixed point in gravity with the existence of minimally coupled matter.
While neglecting matter self-interactions may not be a realistic assumption, 
it is enough to give at least some hint of the effect that matter can have.
We find that within our approximations there are strong
restrictions on the total number of matter fields.
While the standard model seems to be compatible with asymptotic safety of gravity,
many popular scenarios of BSM physics are not.
Our results improve previous work \cite{Percacci:2002ie,Percacci:2003jz}. 
We will discuss  the differences between the new and the old findings in the conclusions.

In section II we will describe  the techniques we employ in some more detail.
In section \red{III} we present the beta functions and the anomalous dimensions of the
graviton, ghost and matter fields, which are the main new technical result of this paper.
Section IV contains the physics results, {\it i.e.}, the bounds on the number of
matter fields and discussion of various models.
We conclude in section V with a summary of the results and cautionary remarks.

\section{Method}
To investigate the coupling of matter to quantum gravity within the asymptotic safety scenario, a method is required that allows for a nonperturbative evaluation of the $\beta$ functions, and an inclusion of (chiral) fermionic, scalar and vector degrees of freedom. Further, a continuum method that allows for analytical calculations is desirable. In the following, we will present such a method.
Here we consider a Riemannian setting, and assume that our results carry over to the Lorentzian case.
\subsection{Functional Renormalization Group}
The functional Renormalization Group is a framework to evaluate $\beta$ functions, even in the nonperturbative regime, where no small coupling exists, and/or nonperturbative threshold effects exist even at small values of the couplings. The method is based upon a Wilsonian momentum-shell wise integration of the path-integral: A mass-like regulator function $R_k(p)$ suppresses quantum fluctuations with momenta $p<k$, where $k$ is an infrared momentum cutoff scale. 
This allows us to define a scale-dependent effective action, the flowing action $\Gamma_k$, which only contains the effect of quantum fluctuations with momenta $p>k$. By changing $k$ we can interpolate smoothly  between the microscopic action $\Gamma_{k \rightarrow \infty}$ and the full quantum effective action $\Gamma_{k \rightarrow 0}$.
The scale-dependence of the flowing action is then given by the Wetterich-equation \cite{Wetterich:1993yh}, a functional differential equation:
\be
\partial_t \Gamma_k = \frac{1}{2} {\rm STr} \left[\left( \Gamma_k^{(2)}+R_k\right)^{-1} \partial_t R_k \right].
\ee
Herein, $\Gamma_k^{(2)}$ denotes the second functional derivative of the flowing action with respect to the fields, and constitutes a matrix in field space. The supertrace $\rm STr$ includes a summation over all discrete indices and the fields, including a negative sign for Grassmann valued fields, i.e., fermions and Faddeev-Popov ghosts in our case. The supertrace also includes a summation over the eigenvalues of the Laplacian in the kinetic term, that translates into a momentum-integral on a flat background. The main technical advantage of the Wetterich equation lies in its one-loop form, which nevertheless takes into account higher-loop effects, as it depends on the full, field-dependent nonperturbative regularized propagator $\left(\Gamma_k^{(2)}+R_k \right)^{-1}$ (see \cite{Papenbrock:1994kf}).
\subsection{Background field and fluctuation field}
In gravity, the background field method is used to set up the RG flow \cite{Abbott:1980hw}. The full metric is split into background $\bar{g}_{\mu\nu}$ and fluctuation $h_{\mu \nu}$:
\be
g_{\mu\nu} = \bar{g}_{\mu\nu}+\sqrt{32\pi G} h_{\mu\nu}.
\ee
We adopt the traditional perturbative convention of rescaling
the metric fluctuation field to make it canonically normalized
(this convention was also used in a functional RG context in \cite{Codello:2013fpa}).
This split allows \red{one} to gauge-fix with respect to the background field, and demanding background-field gauge invariance ensures gauge invariance of the full effective action $\Gamma$. Furthermore, the covariant derivative with respect to the background field allows a meaningful distinction of high- and low-momentum quantum fluctuations, and allows to implement the IR-regularization of the generating functional, while keeping the one-loop form of the Wetterich equation.
Background-independence is achieved by keeping the background field general, and in principle studying all possible backgrounds simultaneously.

The flow equation accordingly depends on two fields, the fluctuation field $h_{\mu \nu}$ and the background field $\bar{g}_{\mu \nu}$. There are two types of terms in which the fluctuation field and the background field do not enter as a sum to combine into the full metric: These are the gauge-fixing term, and the cutoff term. Accordingly, the RG flow will generate terms that depend on the fluctuation metric and the background metric separately. 
This bimetric structure has first been studied in \cite{Manrique:2009uh,Manrique:2010mq,Manrique:2010am}. Here, we extend the recent analysis in \cite{Codello:2013fpa} to include matter fields. 
At the level of the Einstein-Hilbert truncation, we need to introduce the graviton and matter anomalous dimensions, in order to provide a consistent closure of the flow equation from which we will extract the $\beta$ functions of the background field $G$ and $\Lambda$.

Our truncation is given by
\be
\Gamma_k =\Gamma_{\rm EH}+ S_{\rm gf} + S_{\rm gh}+ \Gamma_{\rm matter},
\ee
where the gauge-fixing term is given by:
\bea
S_{\mathrm{gf}}&=& \frac{1}{2\alpha}\int d^d x
\sqrt{\bar g}\, \bar{g}^{\mu \nu}F_{\mu}[\bar{g}, h]F_{\nu}[\bar{g},h]\label{eq:Ggf},\\
 F_{\mu}[\bar{g}, h]&=& \sqrt{2}\, \bar{\kappa} \left(\bar{D}^{\nu}h_{\mu
   \nu}-\frac{1+\rho}{4}\bar{D}_{\mu}h^{\nu}{}_{\nu} \right). 
\eea
Here, $\bar{\kappa}= (32 \pi G)^{-\frac{1}{2}}$ and $\alpha$ and $\rho$ are gauge parameters, which in this paper we choose equal to one.
The standard Faddeev-Popov ghost term in this gauge reads
\bea
 \Gamma_{k\, \rm gh}&=& -\sqrt{2} Z_c  \int d^dx \sqrt{\bg}\,\,\bar{c}_{\mu} 
\Bigl(\bar{D}^{\rho}\bar{g}^{\mu \kappa}g_{\kappa \nu}D_{\rho}\nonumber\\
&+& \bar{D}^{\rho}\bar{g}^{\mu \kappa}g_{\rho \nu}D_{\kappa}
- 
\bar{D}^{\mu}\bar{g}^{\rho \sigma}g_{\rho \nu}D_{\sigma} \Bigr)c^{\nu},
\eea
with a wave-function renormalization $Z_c(k)$, which has first been evaluated in \cite{Groh:2010ta,Eichhorn:2010tb}.
The Einstein-Hilbert term, which depends on the full metric, is given by
\be
\Gamma_{\rm EH}= \frac{1}{16 \pi G} \int d^dx \sqrt{g}\left(-R+2\Lambda \right).
\ee
We then expand the action up to second order in the fluctuation field and introduce a wave-function renormalization for the graviton by $h_{\mu \nu} \rightarrow Z_h^{\frac{1}{2}} h_{\mu \nu}$.
The action then takes the form
\\
\bea
\Gamma_{\rm EH}+ S_{\rm gf}&=&\frac{1}{16 \pi G} \int d^dx \sqrt{\bar{g}}\left(-\bar{R}+2\Lambda \right)
\label{gravity_action}
\\
&{}&
\!\!\!\!\!\!\!\!\!\!\!\!\!\!\!\!\!\!\!\!\!\!\!\!\!\!\!\!\!\!\!\!\!\!\!\!
+\frac{Z_h}{2} \int d^dx \sqrt{\bar{g}} 
\,h_{\mu\nu}K^{\mu\nu\alpha\beta}
((-\bar D^2-2\Lambda)\mathbf{1}^{\rho\sigma}_{\alpha\beta}+W^{\rho\sigma}_{\alpha\beta})
h_{\rho\sigma}\ .
\nonumber
\eea
Here and elsewhere $\mathbf{1}$ is the identity in the space of the fields
(in this instance, symmetric tensors), $K$ is given by
\be
K_{\alpha\beta\rho\sigma}=\frac{1}{2}
\left(\delta_{\alpha \rho}\delta_{\beta\sigma}
+\delta_{\alpha\sigma}\delta_{\beta\rho}
-\delta_{\alpha \beta} \delta_{\rho \sigma}\right)\ ,
\ee
and $W$ is a geometric structure linear in curvature
(see eq.(31) of \cite{Codello:2008vh}) that does not play a role
in the calculation of the anomalous dimensions, that is done on a flat background.
The matter part of the action is given by
\bea
\Gamma_{\rm matter}&=& S_S+S_D+S_V 
\nonumber
\\
S_S&=&\frac{Z_S}{2} \int d^dx \sqrt{g}\,  g^{\mu \nu}\sum_{i=1}^{N_S}  \partial_{\mu} \phi^i \partial_{\nu} \phi^i
\nonumber\\
S_D&=& i Z_D \int d^dx \sqrt{g}\, \sum_{i=1}^{N_D} \bar{\psi}^i \slashed{\nabla} \psi^i,
\nonumber\\
S_V&=&
\frac{Z_V}{4}\int d^d x \sqrt{g}  \sum_{i=1}^{N_V}g^{\mu \nu}g^{\kappa \lambda}F^i_{\mu \kappa}F^i_{\nu \lambda}\nonumber\\
&{}& + \frac{Z_V}{2\xi} \int d^d x \sqrt{\bar{g}}\sum_{i=1}^{N_{\red{V}}}\left(\bar{g}^{\mu \nu}\bar{D}_{\mu}A^i_{\nu}\right)^2
\nonumber\\
&{}& + \frac{1}{2} \int d^d x \sqrt{\bar{g}}\sum_{i=1}^{N_V}\bar C_i(-\bar D^2)C_i\ .
\label{abelian_action}
\eea
In each case, $i$ is a summation index over matter species 
(not to be confused with the representation index of some non-Abelian gauge group).
Similar actions, but without the factors $Z_\Psi$, 
have been considered before in \cite{Dou:1997fg,Percacci:2002ie,Percacci:2003jz}.
Fermions in asymptotically \red{safe} gravity have been further discussed in \cite{Zanusso:2009bs, Vacca:2010mj,Eichhorn:2011pc,Dona:2012am,Harst:2012ni,Gies:2013noa}.
In the Dirac action the covariant derivative is
$\nabla_{\mu}=\partial_{\mu}+\frac{1}{8}[\gamma^{a},\gamma^b]\omega^{ab}_{\mu}$, 
where the spin-connection $\omega^{ab}_{\mu}$ can be expressed in terms of the vielbeins.
This introduces an additional $O(d)$ local gauge invariance.
We adopt a symmetric gauge-fixing of $O(d)$,
such that vielbein fluctuations can be re-expressed completely in terms of the metric fluctuations \cite{Woodard:1984sj,vanNieuwenhuizen:1981uf,Gies:2013noa}.
We will therefore not rewrite the gravitational part of the action in terms of vierbeins;
full details of the procedure we follow can be found in \cite{Dona:2012am}.

There are no gauge interactions, 
so the fermions (as well as the scalars) are uncharged
and there are no gauge covariant derivatives.
In the Abelian gauge field action the second term is a gauge fixing term
with gauge-fixing parameter $\xi$ and the third term represents the abelian ghosts. 
The ghosts are decoupled from the metric and gauge field fluctuations
and therefore do not contribute to the running of $Z_h$ and $Z_V$,
however they are coupled to the gravitational background and
therefore contribute to the beta functions of $G$ and $\Lambda$.
Concerning the question of the mixing between abelian and diffeomorphism 
ghosts as addressed in \cite{Daum:2009dn}, it turns out that there is 
no contribution to the running of $Z_V$, if a regulator is chosen 
that is diagonal in the ghost \red{fields}.
We do not introduce a wave function renormalization for the abelian ghosts in this work.

Our truncation for the gravitational and matter action contains two
essential couplings $G$ and $\Lambda$, 
and five inessential wave-function renormalization constants $Z_\Psi$
with $\Psi=(h,c,S,D,V)$. 
Being inessential means that the $Z_\Psi$ can be eliminated from
the action by field rescalings and do not appear explicitly in any beta function.
As a consequence they are not required to reach a finite limit. 
Instead, when the essential couplings are at their fixed point, they scale 
with a calculable exponent.
It is crucial for our analysis, that the wave-function renormalizations
$Z_\Psi$ nevertheless couple into the beta-functions of the essential couplings 
in a nontrivial way via the anomalous dimensions
\bea
\eta_\Psi &=&-\partial_t\ln Z_\Psi\ .
\eea

The calculation of the beta functions and anomalous dimensions then proceeds as follows.
We choose regulators of type II, in the terminology of \cite{Codello:2008vh}.
For gravitons
\be
\label{gravitoncutoff}
R_{k\, h}(z)=Z_h K\,z\,r\left(\frac{z}{k^2}\right)
\ee
where $z=-\bar D^2+W$. (In the calculation of the anomalous dimensions
one works in flat space and then $z=p^2$).
Likewise for the other fields 
\be
R_{k\, \Psi}(z)=Z_\Psi \mathbf{1}\,z\,r\left(\frac{z}{k^2}\right)\ \ \Psi=c,S,D,V\,
\ee
where $z=-\bar D^2+\mathbf{E}_\Psi$ and $\mathbf{E}_c=-R^\mu_\nu$, 
$\mathbf{E}_S=0$, $\mathbf{E}_D=-R/4$, $\mathbf{E}_V=R^\mu_\nu$.
We use a shape function as in \cite{Litim:2001up}
\be
r\left(\frac{p^2}{k^2} \right)= \left(\frac{k^2}{p^2} -1 \right) \theta(k^2-p^2)\ .
\label{optimized}
\ee
To test the scheme dependence we have also performed calculations with a cutoff of type Ia
on the gravitons, 
which means that $z=-\bar D^2$ in (\ref{gravitoncutoff}).
With these definitions we then evaluate the coefficients of various terms in the r.h.s. of the Wetterich equation.

The beta functions of $G$ and $\Lambda$, and of the corresponding dimensionless couplings 
\be
\tilde{G}=\frac{G}{k^{2-d}}\ ;\qquad
\tilde{\Lambda}= \frac{\Lambda}{k^2}
\ee
are obtained from the coefficients of the terms proportional to $\int d^dx\,\sqrt{\bar g}\bar R$ 
and $\int d^dx\,\sqrt{\bar g}$ in the expansion of the r.h.s. of the Wetterich equation.
They receive contributions from all fields present, including the gravitational and abelian ghosts.

To evaluate the anomalous dimensions $\eta_\Psi$ 
we proceed as in \cite{Eichhorn:2010tb, Vacca:2010mj},
expanding around flat space
and extracting from the r.h.s. of the flow equation terms quadratic in momentum
and in the fluctuation field $\Psi$.
The terms in this calculation have a natural diagrammatic expression
as one loop corrections to the running of the two point function of the field $\Psi$.
The graviton anomalous dimension can be written as
$$
\eta_h=\eta_h\big|_{\rm gravity}+\eta_h\big|_{\rm matter},
$$
where the first term comprises the contributions coming from
graviton and ghost loops, whereas the second comes from matter loops.
The first contribution is evaluated using the diagrams listed in \cite{Codello:2013fpa}.
We project the resulting tensorial structure along the tensor $K$
which is the structure of the graviton propagator in the gauge $\alpha=1$ that
we are using here.

\begin{figure}[!h]
{\resizebox{0.9\columnwidth}{!}
{\includegraphics[width=0.45\linewidth]{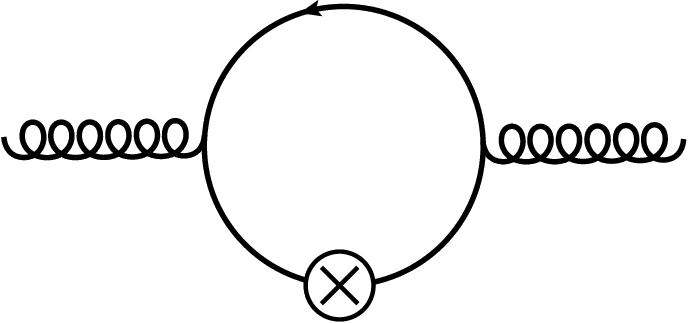}
\qquad
\includegraphics[width=0.45\linewidth]{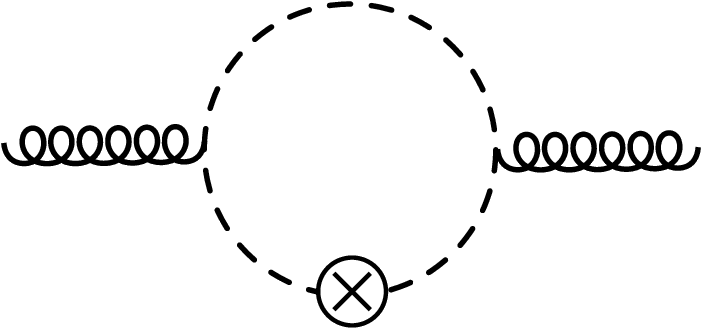}
}
\includegraphics[width=0.45\linewidth]{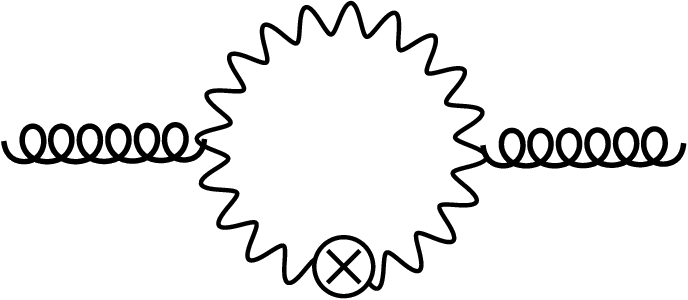}
}
\caption{\label{etah} Diagrams contributing to $\eta_{h}$. Spirals denote gravitons, thick lines fermions, dashed lines scalars and curly lines gauge fields. 
A crossed circle denotes an insertion of  $\partial_t R_k$.}
\end{figure}

The matter contribution to $\eta_h$ is given by the diagrams of fig.~\ref{etah},
again projected along $K$.
In principle, tadpole diagrams exist which could also contribute to $\eta_h$. 
The momentum structure of 
the vertices in our truncation implies a vanishing of those diagrams.

As we neglect matter-ghost couplings at this order of the approximation 
\cite{Eichhorn:2013ug}, the ghost anomalous dimension will be the same 
as in the pure gravity case, and there is no ghost contribution to the matter anomalous dimensions.

\begin{figure}[!h]
\begin{minipage}{0.5\linewidth}
\includegraphics[width=\linewidth]{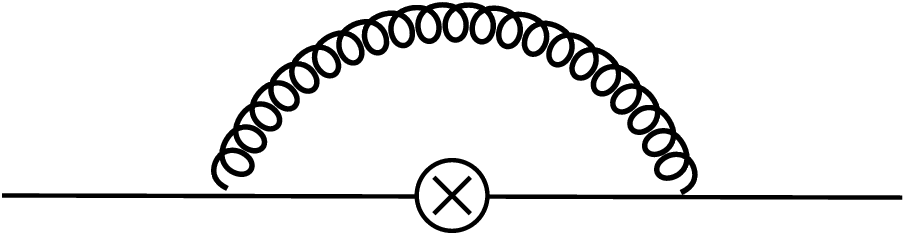} \\
\vspace{0.5cm}
\includegraphics[width=\linewidth]{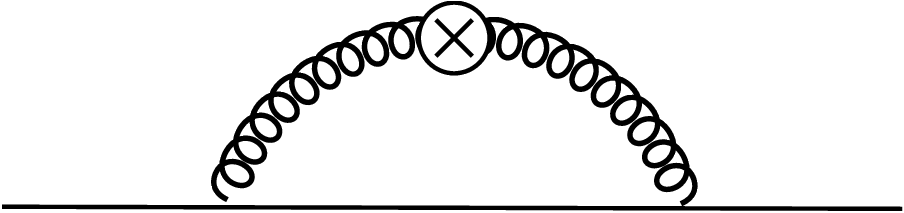}
\end{minipage}
\begin{minipage}{0.3\linewidth}
\includegraphics[width=0.8\linewidth]{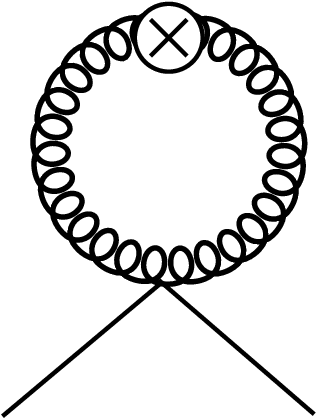}
\end{minipage}
\caption{\label{etapsi} Diagrams contributing to $\eta_D$. 
Analogous diagrams with external scalar and vector fields determine $\eta_S$ and $\eta_V$.}
\end{figure}

Finally, graviton fluctuations induce nontrivial matter anomalous dimensions
through the diagrams of~fig.~\ref{etapsi}.

\FloatBarrier

\begin{widetext}

\section{Beta functions}
The beta functions for $\tilde{G}$ and $\tilde{\Lambda}$ have the following form:
\begin{eqnarray}
\label{generallambda}
\frac{d\tilde\Lambda}{dt}&=&\!
-2\tilde\Lambda
+\frac{8\pi\tilde{G}}{(4\pi)^{d/2}d(d+2)\Gamma\red{(}d/2\red{)}}
\Biggl[\frac{d(d+1)(d+2-\eta_h)}{1-2\tilde\Lambda}
-4d(d+2-\eta_c)
\nonumber
\\
&{}&
\qquad\qquad\qquad
+2N_S(2+d-\eta_S)
-2N_D 2^{[d/2]}(2+d-\eta_D)
+2N_V(d^2-4-d\,\eta_V)\Biggr]
\nonumber\\
&{}& 
\qquad
-\frac{4\pi\tilde{G}\tilde\Lambda}{3d (4\pi)^{d/2}\Gamma\red{(}d/2\red{)}}
\Biggl[\frac{d(5d-7)(d-\eta_h)}{1-2\tilde\Lambda}
+4(d+6)(d-\eta_c)
\nonumber\\
&{}&
\qquad\qquad\qquad
-2N_S(d-\eta_S)
-N_D 2^{[d/2]}(d-\eta _D)
+2N_V(d\,(8-d)-(6-d)\eta_V)\Biggr]
\\
\label{generalg}
\frac{d\tilde G}{dt}&=&\!\!
(d-2)\tilde G
-\frac{4\pi\tilde{G}^2}{3d(4\pi)^{d/2}\Gamma(d/2)}
\red{\Biggl[}\frac{d(5d-7)(d-\eta_h)}{1-2\tilde\Lambda}
+4(d+6)(d-\eta_c)
\nonumber\\
&{}&
\qquad\qquad\qquad
-2N_S(d-\eta_S)
-N_D 2^{[d/2]}(d-\eta _D)
+2N_V(d(8-d)-(6-d)\eta_V)\Biggr]\ .
\end{eqnarray}
Except for the appearance of the matter anomalous dimensions,
this agrees with \cite{Codello:2008vh}.
The main novel computational \red{results} of this paper are the 
formulas for the anomalous dimensions 
$\eta_h$, $\eta_S$, $\eta_D$ and $\eta_V$. 
Following \cite{Codello:2013fpa}, the gravitational contribution 
to the graviton anomalous dimension can be written in the form
\be
\label{results_etah1}
\eta_h\Big|_{\rm gravity}=\left[a(\tilde\Lambda_k)
+c(\tilde\Lambda_k)\eta_h
+e(\tilde\Lambda_k)\eta_c\right]\tilde G_k\ .
\ee
We have checked the results of \cite{Codello:2013fpa} when $\eta_h$ is defined
by projecting on the spin two propagator. 
With our definition of $\eta_h$, which involves projection on the tensor $K$
the coefficients turn out to be:
\bea
a(\tilde\Lambda)&=&\frac{a_0+a_1\tilde\Lambda+a_2\tilde\Lambda^2
+a_3\tilde\Lambda^3+a_4\tilde\Lambda^4}{(4\pi)^{d/2}\Gamma(d/2)d^2(d^2-4)(3d-2)(1-2\tilde\Lambda)^4}\ ,
\\
a_0&=&-4\pi\left(d-2\right)(-896+264\, d+1076\,  d^2-434 \, d^3+21\,  d^4+d^5)\ ,
\nonumber
\\
a_1&=&16\pi\left(d-2\right)(-2048+2552\,  d-318 \, d^2-125 \, d^3+2\,  d^4+d^5)\ ,
\nonumber
\\
a_2&=&-16\pi(12544-25760\,  d+16968 \, d^2-4228 \, d^3+354 \, d^4-17 \, d^5+d^6)\ ,
\nonumber
\\
a_3&=&4096\pi(d-2)(-32+50\,  d-19\,  d^2+2 \, d^3)\ ,
\nonumber
\\
a_4&=&-2048\pi(d-2)(-32+50 \, d-19\,  d^2+2\,  d^3)\ ;
\nonumber\\
c(\tilde\Lambda)&=&\frac{8\pi(d-1)\left[128+720\,  d-350 \, d^2+29 \, d^3+32(d-2)(d+4)\tilde\Lambda\right]}{(4\pi)^{d/2}\Gamma(d/2)\,d^2(d+2)(d+4)(3d-2)(1-2\tilde\Lambda)^3}
\\
e(\tilde\Lambda)&=&-\frac{128\pi\left(32-50\, d+23\, d^2\right)}{(4\pi)^{d/2}\Gamma(d/2)d^2(d+2)(d+4)(3\, d-2)}\ .
\eea
The matter contribution is
\bea
\label{results_etah3}
\eta_h\Big|_{\rm matter}&=&
-N_S\,\frac{32\pi\tilde G}{(4\pi)^{d/2}\Gamma(d/2)}\,\frac{1}{d^2(d+2)(3d-2)}
\left[(d-2)^3+2\frac{8-10\, d+d^2}{d+4}\,\eta_S
\nonumber
\right]\\
&& +N_D\, 2^{[d/2]}\frac{16\pi\tilde{G}}{(4\pi)^{d/2}\Gamma(d/2)}
\,\frac{(d-1)(d-2)}{d^{3}(3d-2)}\left[2+\frac{d-2}{d+1}\,\eta_D\right]\\
&& -N_V\,\frac{32\pi\tilde{G}}{(4\pi)^{d/2}\Gamma(d/2)} \frac{(d-1)(d-2)}{d^2(d+2)(3d-2)} \left[d^2-12\, d+8-2\frac{16-d}{d+4}\eta_V\right],
\nonumber
\eea
The ghost anomalous dimension is
\be
\label{results_etah2}
\eta_c=\left[b(\tilde\Lambda_k)
+d(\tilde\Lambda_k)\eta_h
+f(\tilde\Lambda_k)\eta_c\right]\tilde G_k\ .
\ee
with
\bea
b(\tilde\Lambda)&=&\red{\frac{64\pi\left[-8+4\, d+18\, d^2-7\, d^3+2(4-9\, d^2+3 \, d^3)\tilde\Lambda\right]}{(4\pi)^{d/2}\Gamma(d/2)d^2(d^2-4)(1-2\tilde\Lambda)^2}}
\\
d(\tilde\Lambda)&=&\red{\frac{64\pi(4-4\,  d-9 \, d^2+4 \, d^3)}{(4\pi)^{d/2}\Gamma(d/2)d^2(d^2-4)(d+4)(1-2\tilde\Lambda)^2}}
\\
f(\tilde\Lambda)&=&\red{\frac{64\pi(4-9 \, d^2+3\,  d^3) }{(4\pi)^{d/2}\Gamma(d/2)d^2(d^2-4)(d+4)(1-2\tilde\Lambda)}}
\eea
Finally the anomalous dimensions of the matter fields are
\bea
\eta_S&=&-\frac{32\pi\tilde{G}}{(4\pi)^{d/2}\Gamma(d/2)}
\Biggl[\frac{2}{d+2}\frac{1}{(1-2\tilde{\Lambda})^2}\,\left(1-\frac{\eta_{h}}{d+4}\right)
+\frac{2}{d+2}\frac{1}{1-2\tilde{\Lambda}}\left(1-\frac{\eta_S}{d+4}\right)\nonumber\\
&{}& 
\phantom{- 32\pi g\frac{2\pi^{d/2}}{2(2\pi)^{d}\Gamma(d/2)}}
+\frac{(d+1)(d-4)}{2d(1-2\tilde{\Lambda})^2}\left(1-\frac{\eta_{h}}{d+2}\right)\Biggr],\\
\label{results_etas}
\eta_D&=&\frac{32\pi\tilde{G}}{(4\pi)^{d/2}\Gamma(d/2)}  \Biggl[\frac{(d-1)(d^2+9\, d-8)}{8d\, (d-2)(d+1)(1-2\tilde\Lambda)^2}\left(1-\frac{\eta_{h}}{d+3}\right)
\nonumber
\\
&{}&
\qquad\qquad
+\frac{(d-1)^{2}}{2d(d+1)(d-2)}\frac{1}{1-2\tilde{\Lambda}}
\left(1-\frac{\eta_D}{d+2}\right)
-\frac{(d-1)(2d^2-3d-4)}{4d(d-2)(1-2\tilde{\Lambda})^2}\left(1-\frac{\eta_h}{d+2}\right)\Biggr] \\
\label{results_etaD}
\eta_V&=&-\frac{32\pi\tilde{G}}{(4\pi)^{d/2}\Gamma(d/2)}
\Biggl[\frac{(d-1)(16+10\, d-9\, d^2+d^3)}{2d^2(d-2)(1-2 \tilde{\Lambda})^2} \left(1-\frac{\eta_h}{d+2}\right) 
\nonumber
\\
&{}&
\qquad\qquad
+\frac{4(d-1)(2d-5)}{d(d^2-4)(1-2 \tilde{\Lambda})}\left(1-\frac{\eta_V}{d+4}\right)
+\frac{4(d-1)(2d-5)}{d(d^2-4)(1-2 \tilde{\Lambda})^2}\left(1-\frac{\eta_h}{d+4}\right)\Biggr]
\label{results_etaV}.
\eea
\end{widetext}

\goodbreak

\section{Results}

\subsection{Perturbative approximation}

\begin{figure}[!h]
\includegraphics[width=\linewidth]{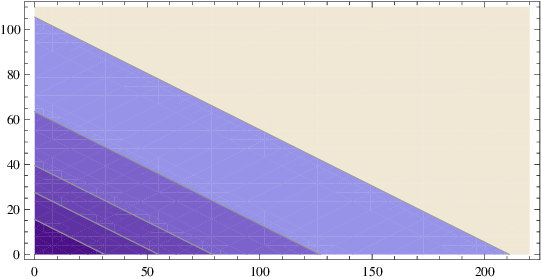}
\caption{\label{1loopPlot}The shaded triangles are the areas in the $N_S$-$N_D$ 
plane compatible with a gravitational fixed point 
for $N_V=0$ (darkest triangle in bottom left corner),
$N_V=6,12,24,45$ (from bottom to top).}
\end{figure}

In order to get a rough idea of the effect of matter on the RG flow,
in a context where solutions can be found analytically rather than numerically,
it is useful to consider first the perturbative approximation, 
which consists of neglecting all anomalous dimensions and expanding
the beta functions to second order in $\tilde\Lambda$ and $\tilde G$.
This is justified in some neighborhood of the Gaussian fixed point.
In $d=4$ the beta functions become
\bea
\beta_{\tilde G} = 2\tilde G + \frac{\tilde G^2}{6 \pi }\left(N_S+2 N_D-4 N_V-46\right),\\
\label{oneloop}
\beta_{\tilde \Lambda} = -2\tilde \Lambda +\frac{\tilde G}{4\pi}\left(N_S-4 N_D+2N_V+2\right)
\nonumber\\
+\frac{\tilde G \tilde \Lambda }{6\pi}\left(N_S+2N_D-4 N_V-16\right).
\eea
The numbers $46$, $2$ and $16$ represent the contributions of gravitons and ghosts
to the beta functions.
These contributions are such that the RG flow admits a nontrivial fixed point
when matter is absent.
Let us see what effect matter has, in this approximation.
The beta \red{functions} have a nontrivial fixed point at
\bea
\label{lfp}
\tilde\Lambda_*&=&-\frac{3}{4}\frac{N_S-4N_D+2 N_V+2}{N_S+2N_D-4N_V-31}\ ,
\\
\label{nfp}
\tilde G_*&=&-\frac{12\pi}{N_S+2N_D-4N_V-46}\ .
\eea
Since the beta functions vanish for $\tilde G=0$, flow lines cannot cross 
from negative to positive $\tilde G$.
Since the low energy Newton's coupling is experimentally bound to be positive,
we require that also the fixed point occurs at positive $\tilde G$.
This puts a bound on the matter content.
In the following we shall find it convenient to present the results
in the $N_S$-$N_D$-plane, treating the number of gauge fields as a fixed parameter.
Positivity of $\tilde G_*$ demands that
\be
\label{gsing}
N_D<23+2N_V-\frac{1}{2}N_S\ .
\ee
Notice that gauge fields contribute with the same sign as gravity,
so they facilitate the existence of the fixed point, whereas scalars and fermions tend to destroy it.
When their number increases, the fixed-point value of $\tilde G_*$ increases
and reaches a singularity on the line $N_D=\red{23}+2N_V-\frac{1}{2}N_S$.
On the other side of the singularity $\tilde G_*$ is negative.
Fig.~\ref{1loopPlot} shows the existence region of a positive fixed point for $\tilde G_{\ast}$
for no gauge fields or $12$, $24$, $45$ gauge fields.
(The significance of these numbers will be discussed later.)
We see that the existence region grows with the number of gauge fields,
but most importantly, for a given number of gauge fields,
only a finite number of combinations of scalar and Dirac fields is allowed.

The cosmological constant has a singularity on the line 
\be
\label{lsing}
N_D=\frac{31}{2}+2N_V-\frac{1}{2}N_S.
\ee
This singularity in $\tilde\Lambda_*$ is parallel to the singularity in $\tilde G_*$
and is shifted downwards by $N_D=7.5$.
There are fixed points in the intermediate region between
these singularities, but they are disconnected from the one in the origin
({\it i.e.}, with no matter), so we regard them as physically very untrustworthy.
For ``phenomenological'' applications we will restrict our attention to
points that are below the singularity in the cosmological constant.
The allowed region is therefore somewhat smaller than the one
shown in fig.~\ref{1loopPlot}.
In the absence of gauge fields this leaves only the area
$N_D<\frac{31}{2}-\frac{1}{2}N_S$,
which means that at most \red{30} Weyl spinors or \red{30} scalars,
or a \red{combination} thereof, are admissible. 

When we restrict ourselves to the allowed region,
the sign of the cosmological constant at the fixed point is determined 
by the numerator in (\ref{lfp}):
above (or left) of the line 
\be
N_D=\frac{2+2 N_V+N_S}{4}
\label{susy}
\ee
the cosmological constant is negative, 
whereas below (or right) of this line it is positive.
Note that in the beta function for $\tilde\Lambda$ the contribution of each field
is \red{weighted} with the number of degrees of freedom it carries,
with a plus sign for bosons and a minus sign for fermions.
The line (\ref{susy}) is where any supersymmetric theory would lie.
The contours of constant $\tilde\Lambda_*$ are straight lines passing through 
the point $(2N_V+20,N_V+11/2)$, where (\ref{lsing}) and (\ref{susy}) intersect.
The singularity of the cosmological constant on \eqref{lsing}
is negative left of this point and positive right of it.

The stability matrix
\be
M=\left(
\begin{array}{cc}
\frac{\partial\beta_{\tilde\Lambda}}{\partial\tilde\Lambda}
&\frac{\partial\beta_{\tilde\Lambda}}{\partial\tilde G}\\
\frac{\partial\beta_{\tilde G}}{\partial\tilde\Lambda}
&\frac{\partial\beta_{\tilde G}}{\partial\tilde G}\\
\end{array}
\right)
\ee
has eigenvalues $-2$ and
$-4\frac{N_S+2N_D-4N_V-31}{N_S+2N_D-4N_V-46}$.
Below the singularites of $\tilde\Lambda_*$ and $\tilde G_*$,
the numerator and denominator of this ratio are positive,
so both eigenvalues are negative.
In the region between the singularities the second eigenvalue would be positive.

In this perturbative
approximation one can examine \red{the} effect of matter on higher gravitational couplings.
If we parametrize the curvature squared terms, up to total derivatives, as
\begin{equation}
\label{actionansatz}
\int d^4x\,\sqrt{g}\left[
\frac{1}{2\lambda}C^{2}+\frac{1}{\xi}R^{2}
\right],\,
\end{equation}
where $C$ is the Weyl tensor, the beta functions of the couplings are
\begin{align}
\beta_{\lambda} & =  -\frac{1}{(4\pi)^{2}}\frac{133}{10}\lambda^{2}-2\lambda^2 a^{(4)}_\lambda\ ,\nonumber\\
\beta_{\xi} & =  -\frac{1}{(4\pi)^{2}}\left(10\lambda^2-5\lambda\xi+\frac{5}{36}\xi^2\right)-\xi^2 a^{(4)}_\xi\
,\nonumber
\end{align}
where
\begin{align}
\label{bsq}
a^{(4)}_\lambda&= \frac{1}{2880\pi^2}\left(\frac{3}{2}N_S+9N_D+18N_V\right)\ ,\\
a^{(4)}_\xi&= \frac{1}{2880\pi^2}\frac{5}{2}N_S\ .
\end{align}
It is remarkable that all types of matter contribute with the same sign
to the running of these couplings, which are always asymptotically free.

In the perturbative approximation it is easy to compute the beta functions also with other definitions of the cutoff.
If we choose the so-called type Ia cutoff on gravitons (see \cite{Codello:2008vh})
the one loop beta functions are given by
\bea
\beta_{\tilde{G}}
&=& 2\tilde{G} + \frac{\tilde{G}^2}{6 \pi }\left(N_S+2 N_D-4 N_V-22\right),\\
\label{oneloop}
\beta_{\tilde{\Lambda}}
&=& -2\tilde{\Lambda}+\frac{\tilde{G}}{4\pi}\left(N_S-4 N_D+2N_V+2\right)
\nonumber
\\
&&+\frac{\tilde{G}\tilde{\Lambda}}{6\pi}\left(N_S+2N_D-4N_V+8\right).
\eea
Notice that the matter contribution has not changed:
for massless scalars the two types of cutoff are the same,
for fermions we must always use the type II cutoff \cite{Dona:2012am}
and for simplicity we have maintained this cutoff also for gauge fields.
The only difference is therefore in the gravitational contribution.
One can repeat the preceding discussion with little changes.
The main effect is that the permitted region is smaller, with the
singularities shifted downwards:
23 is replaced by 11 in (\ref{gsing}) and 31/2 is replaced by 7/2 in (\ref{lsing}).
We can view these shifts as a measure of the 
typical theoretical uncertainties in this approximation.

The coefficients (\ref{bsq}) are universal 
and, as noticed in \cite{Percacci:2005wu}, with type II cutoff and with the
shape function (\ref{optimized}) the contribution of matter to the running
of all couplings multiplying terms with six or more derivatives is identically zero.

\goodbreak

\subsection{The full system}

\subsubsection{Anomalous dimensions and RG improvement}

Next we want to analyze the fixed point of the full nonlinear system
of beta functions (\ref{generallambda},\ref{generalg}),
including the anomalous dimensions.
The formulas (\ref{results_etah1}-\ref{results_etaV}) do not directly give
the anomalous dimensions, rather they give a set of linear equations 
for the anomalous dimensions.
The appearance of the anomalous dimensions on the r.h.s. of these equations is due to the fact that couplings that enter the regulator function (in this case, the wave function renormalizations $Z_{\Psi}$) have to be treated as running parameters.
If we denote $\vec\eta=(\eta_h,\eta_c,\eta_S,\eta_D,\eta_V)$,
these equations can be written in the form
\be
\vec\eta=\vec\eta_1(\tilde\Lambda,\tilde G)+A(\tilde\Lambda,\tilde G)\vec\eta\ .
\ee
where $\vec\eta_1$ is the leading one loop term and $A$ is a matrix of coefficients.

The reason for calling $\eta_1$ the one loop anomalous dimension is that in the functional RG
the one loop approximation consists precisely of neglecting the running of the couplings
in the r.h.s. of the Wetterich equation.
To avoid misunderstandings, let us also comment that in a single-field truncation,
where one neglects the terms of order $h$ in (\ref{gravity_action}),
the anomalous dimension is identified with 
\be
\label{gloria}
-k\frac{d}{dk}\log\left(\frac{1}{16\pi G}\right)\ .
\ee
This is the origin of the often-made statement that the one loop approximation
consists of neglecting the anomalous dimensions.
In a two-field truncation, with independent wave function renormalization for the
fluctuation field, this statement is not true and one can have anomalous dimensions
at one loop.

In order to write the anomalous dimensions as functions of $\tilde G$ and $\tilde\Lambda$
one has to solve this system of equations.
We refer to this as ``the RG improvement''.
The resulting expressions are considerably more complicated than the ones
appearing in eq. (\ref{results_etah1}-\ref{results_etaV}).
In particular they are rational functions in $\tilde\Lambda$ and $\tilde G$
whose numerators and denominators are polynomials of higher order than the 
ones appearing in the leading one loop terms.
Since the full equations contain polynomials of higher order than the
leading ones, they will also have more solutions.
In general we consider to be reliable those features 
of the system that can be seen already in simple approximations and that persist 
when more complicated features are taken into account.
This implies that all the additional solutions of the full
system that are not present in the one loop system are spurious.

Furthermore, in situations where the anomalous dimensions become large,
the improvement terms can become numerically dominant relative
to the leading terms, in which case also the true solutions may exhibit features
that are non-physical.
Clearly this means that the ``RG improved'' results have to be taken with
great care when the anomalous dimensions become large.
In order to avoid potential pitfalls due to these facts, unless otherwise stated
in what follows we shall present the results taking only the leading terms of the anomalous
dimensions into account.
We will discuss explicitly some cases when the full nonlinear system can be studied
and gives reliable results.

\subsubsection{Selection criteria}

Even the leading one loop flow equations are very nonlinear,
and for any given triple $(N_D, N_S, N_V)$, 
there may be several fixed points.
How do we know whether a fixed point is physically significant
or just an artifact of the truncation?
Since the nontrivial fixed point in the absence of matter is relatively well understood,
we try to select among all possible fixed points in the presence of matter the one that
derives from a continuous deformation of the fixed point without matter.
The following criteria are also useful.

\begin{itemize}
\item We discard those fixed points for which $\tilde{G}_{\ast}<0$. As already remarked,
although the fixed-point value of $\tilde{G}$ is not restricted by observations, its low-energy value is. Thus a realistic model of gravity must show an RG flow towards the IR, such that $\tilde{G}(k_{\rm IR})>0$. To the best of our knowledge, no truncation exists in which $\tilde{G}$ changes sign under the RG flow, thus ruling out $\tilde{G}_{\ast}<0$.
\item We discard fixed points which have less than two relevant directions. While in principle the low-energy value of the Newton coupling or the cosmological constant could be a prediction of the theory, both correspond to free parameters of the pure-gravity theory. We expect that for a small number of matter degrees of freedom, the number of critical exponents should not change, otherwise the truncation would be insufficient. This does not rule out the possibility that a very large number of matter degrees of freedom leads to substantial changes in the properties of the theory and a viable fixed point has only one relevant direction, but we do not consider this possibility in the following.
\end{itemize}
Following this procedure, we find severe restrictions on the number of matter and gauge fields compatible with asymptotically safe gravity. 
Note that some of the fixed points found in this way can have rather large critical exponents. Such large values indicate a huge departure from canonical scaling and imply that quantum fluctuations have a very big effect. Thus our truncation is presumably insufficient to capture the relevant physics in this case, and yields unreliable results.

\subsubsection{Anomalous dimensions and predictivity}

A connection exists between the anomalous dimension of the fields and the critical exponents at an interacting fixed point, so that we can deduce a bound on the anomalous dimension by requiring predictivity of the theory: Let us consider an operator $\mathcal{O}= \Phi^n$, where $\Phi$ stands for any of the fluctuation fields of the theory, e.g., the graviton. The dimensionality of the corresponding coupling $g_{\mathcal{O}}$ is then given by $d_{g}= d-n\, d_{\Phi}$, where $d_{\Phi}$ is the dimensionality of the field. Accordingly, the dimensionless coupling $\tilde{g}_{\mathcal{O}}$ is given by
\be 
\tilde{g}_{\mathcal{O}}= g_{\mathcal{O}} \frac{k^{-d+n d_{\Phi}}}{Z_{\Phi}^{\frac{n}{2}}}.
\ee
The $\beta$ function for the coupling $\tilde{g}_{\mathcal{O}}$ will thus have the following structure
\be
\beta_{\tilde{g}_{\mathcal{O}}} = \left(-d + n d_{\Phi}+ \frac{n}{2} \eta_{\Phi} \right)\tilde{g}_{\mathcal{O}}+\dots\, \, ,
\ee
where we have introduced the anomalous dimension $\eta_{\Phi}= - \partial_t \ln Z_{\Phi}$. The additional terms in the $\beta$ function depend on the particular operator that we consider, and are nonzero at an interacting fixed point. Neglecting operator mixing,
they will result in a shift of the critical exponent $\theta_{\mathcal{O}}$ 
from the canonical value (which it has at a noninteracting fixed point) to
\be
\theta_{\mathcal{O}} = - \frac{\partial \beta_{\tilde{g}_{\mathcal{O}}}}{\partial \tilde{g}_{\mathcal{O}}} \vert_{\tilde{g}_{\mathcal{O}}= \tilde{g}_{\mathcal{O} \ast}} = d-n d_{\Phi}-\frac{n}{2}\eta_{\Phi}+ \dots\, \, .
\ee 
The sign of the critical exponent cannot be determined from general arguments, but must be fixed by an explicit calculation.  
At an interacting fixed point, the anomalous dimension constitutes a further departure from canonical scaling, that scales with $n$. 
Predictivity demands that at most a finite number of operators 
should be shifted into relevance at an interacting fixed point.
This implies that
\be
\eta_{\Phi}> -2 d_{\Phi} + \frac{2 d}{n} \xrightarrow{n \rightarrow \infty} -2 d_{\Phi}.
\ee
In the case of the graviton, $d_h=\frac{d-2}{2}$, therefore
\be
\eta_h > -d+2.
\ee
Considering operators ${\cal O}$ that contain derivatives will generally
give a weaker bound; here we will consider the strongest possible bound.
We will use this as a fourth criterion to bound the number of matter fields compatible with asymptotic safety.
We will not take into account similar bounds on the matter anomalous dimension, as we have neglected all matter self-interaction. We thus assume that our values for the matter anomalous dimensions will change in a more complete truncation.

In the following, we apply the criteria specified above and discuss the compatibility of scalars, Dirac fermions and vectors with a viable fixed point for gravity.

\subsection{Fixed points}

\subsubsection{No matter}

The anomalous dimensions of the ghosts had been calculated
previously in \cite{Eichhorn:2010tb}, \cite{Groh:2010ta}.
The running of the graviton \red{two-point function} had been calculated
in \cite{Donkin:2012ud}, \cite{Christiansen:2012rx},
where it was interpreted as running of the cosmological and Newton couplings.
Here we follow \cite{Codello:2013fpa} in resolving the difference between the running of the 
Newton coupling and the anomalous dimension.
At this stage, we ignore the difference between the running of the background 
cosmological constant and the mass term in the graviton propagator.
Our derivation differs from previous ones in the definition of the cutoff
and of the anomalous dimension.
We use a type II cutoff, in part for coherence with the cutoff in the fermionic sector,
but also because it leads to beta functions containing polynomials of lower order 
and hence with fewer spurious solutions.
Furthermore, in the definition of the anomalous dimension $\eta$
we have projected the two-point function on the tensor $K$,
which is the structure it has in the internal lines.
This also has the computational advantage of depending only on the metric and not on
the external momenta.
We list the results in the following table.

\begin{table}[!h]
\caption{}
\begin{ruledtabular}\label{mattermodeltab} 
\begin{tabular}{ccccc}
 & 1L-II & full-II & full-Ia &Ref. \cite{Codello:2013fpa} \\\hline
\ $\tilde\Lambda_*$ & $0.010$ & $0.009$ & $-0.049$\ &$-0.008$\ \\\hline
\ $\tilde G_*$ & $0.772$ & $0.776$ & $1.579$ & $1.446$ \\\hline
\ $\theta_1$ & $3.298$ & $3.317$ & $3.991$&$3.323$\\\hline
\ $\theta_2$ & $1.954$ & $1.925$ & $1.290$&$1.954$\\\hline
\ $\eta_h$& $0.269$ & $0.299$& $0.540$&$0.072$\\\hline
\ $\eta_c$ & $-0.806$ &  $-0.814$&$-1.390$& $-1.503$
\end{tabular}
\end{ruledtabular}
\end{table}

The first two columns give the results of the one loop approximation,
as defined in subsection IV.B.1, and the RG improved equations,
both with the type II cutoff.
The difference is very small, in accordance with the fact that
the anomalous dimension of the graviton is small.
The third column gives the result we obtain using a cutoff of type Ia
instead of II.
The difference with the first column is not very small quantitatively,
in line with previous discussions of the scheme-dependence in this approach
\cite{Reuter:2001ag,Narain:2009qa}.
A higher order truncation would be needed to improve this aspect.
The last column gives the results of reference \cite{Codello:2013fpa},
who also used a cutoff of type Ia.
The differences that are seen between the last two columns can thus
all be attributed to our different definition of the anomalous dimensions.
The main difference between these results and the earlier literature 
lies in the real critical exponents,
which are seen to be only weakly dependent on the technical details.
We anticipate that this is mainly due to our identification of the
``graviton mass'' (the term quadratic in $h$ and without derivatives in (\ref{gravity_action}))
with the background cosmological constant. When the graviton mass is allowed to run
independently the flow in the $\tilde\Lambda$-$\tilde G$ plane has again
complex critical exponents, while the flow in the mass-$\tilde G$ plane
looks like the one discussed here.
We will return to this point in a future publication.
\smallskip

\subsubsection{Scalar matter}

Even though physically $N_S$ must be an integer, mathematically one can
study the dependence treating $N_S$ as a continuous parameter.
For $N_S\leq 12$ the effect of scalars is to push $\tilde{\Lambda}_*$ towards larger values,
while $\tilde G_*$ is almost stable.
The product $\tilde{\Lambda}_*\tilde G_*$, which is generally known to be
quite independent of technical details such as gauge and cutoff choice,
increases slowly, see fig.~\ref{GLambdaNS}. 
In this regime the critical exponents change little 
while the anomalous dimensions increase in absolute value,
maintaining the same sign ($\eta_h>0$, $\eta_c<0$ and $\eta_S<0$).
There is a sharp change of behavior of $\tilde\Lambda_*$ for $N_S\geq 12$.
Beyond this value, the cosmological constant stops growing with $N_S$,
while $\tilde{G}_{\ast}$ begins to grow and also the critical exponents become very large ($O(10^3)$).

\begin{widetext}

\begin{figure}[!h]
{\resizebox{1\columnwidth}{!}
{\includegraphics{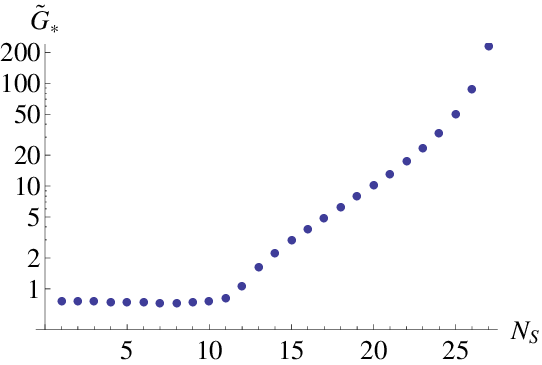}
\includegraphics{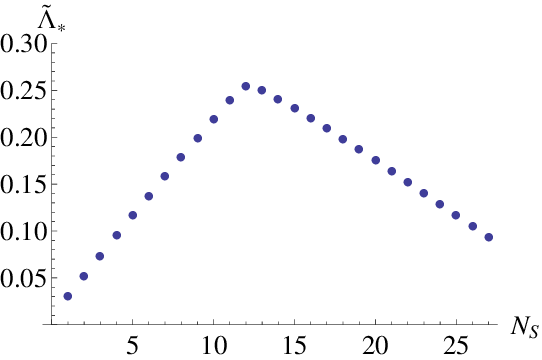}
\includegraphics{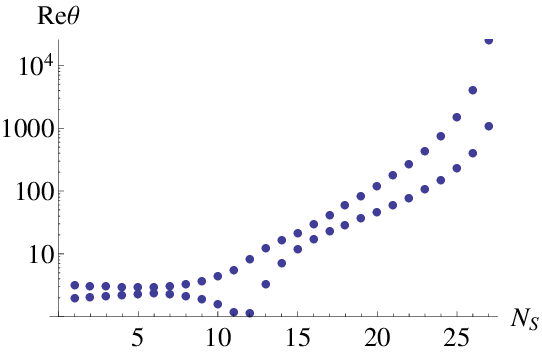}}
\caption{\label{GLambdaNS} Left and middle: Position of the fixed point as a function of the number of scalar fields. Right: critical exponents. Note the logarithmic scales.
All with type II cutoff and one loop anomalous dimensions.}
}
\end{figure}

\begin{figure}[!h]
{\resizebox{1\columnwidth}{!}
{\includegraphics{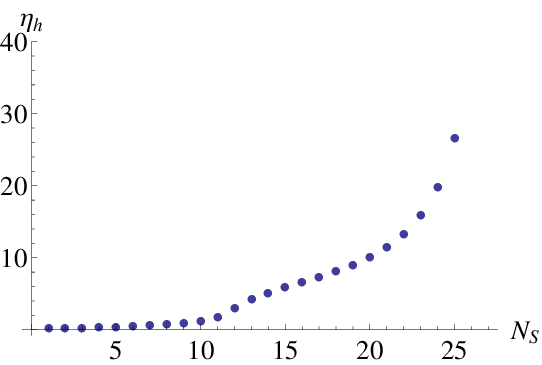}
\includegraphics{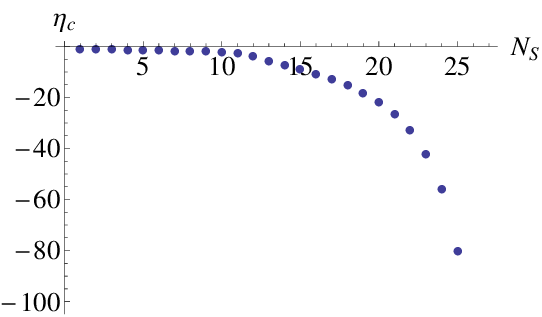}
\includegraphics{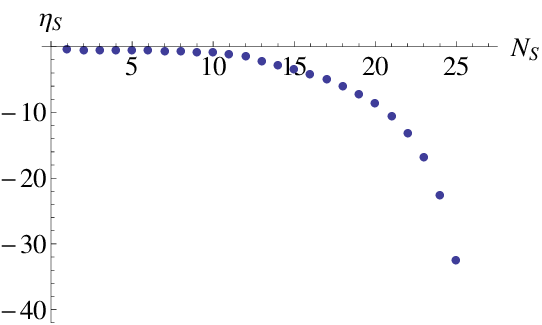}
}
\caption{\label{etaNS} The graviton (left), ghost (middle) and scalar (right) anomalous dimensions as functions of the number of scalar fields.}
}
\end{figure}

\end{widetext}

As far as we could see, the change of behavior occurs smoothly 
over the whole range, so one has a continuous deformation of the
pure gravity fixed point up to $N_S\approx27.7$ where $\tilde G_*$ diverges.
As in the perturbative approximation, there is therefore a maximal number of scalar fields 
that is compatible with the existence of a viable fixed point.
This is in contrast to \cite{Percacci:2002ie,Percacci:2003jz},
where the fixed point seemed to exist for any number of scalars.
We believe that fixed point to be an artifact of the identification
of $\eta_h$ with (\ref{gloria}) in the single field truncation.

The effect of scalar fields on the position of the fixed point,
on the critical exponents and on the anomalous dimensions is shown
in figs.~\ref{GLambdaNS},\ref{etaNS}, at one loop and with type II cutoff.
Including the RG improvement results in more complicated behaviour.
While the fixed-point value for the Newton coupling is nearly constant up to $N_S=11$, 
it rises sharply thereafter.
At the same time, $\tilde\Lambda_*\approx0.25$ becomes nearly independent of $N_S$.
The critical exponents are complex for $12\leq N_S\leq50$ and  
their real part is negative for $15 \leq N_S \mathrel{\red{\leq}} 20$.
The singularity is deferred to $N_S\approx 85$.

With a type Ia cutoff the anomalous dimensions remain smaller
and the fixed point becomes complex at $N_S\approx 17$.
This lower limit is in line with the result of the perturbative approximation.
We thus observe a significant scheme-dependence for $N_S>12$.
This, together with the fact that the anomalous dimensions 
become rather large in that range, makes the full RG improved equations unreliable.
This suggests, that the fixed point beyond $N_S=17$ could be a truncation artifact.

In the future we may understand better which truncation gives physically reliable results
but for the time being the scheme dependence has to be taken as a measure of the theoretical
uncertainties.
For now we can say with good confidence that the fixed point ceases to exist
when the number of scalars becomes of the order of $22\pm5$.
To sharpen this number, one should study the behavior of higher background curvature terms,
for example repeating the analysis in \cite{Falls:2013bv} under the inclusion of scalars.
On the other hand, in order to understand whether the large negative scalar anomalous dimension
could lead to an increase in the number of relevant directions, one should
study this question in the presence of scalar self-interactions \cite{Eichhorn:2012va}.

\smallskip

\subsubsection{Fermionic matter}

\begin{figure}[!h]
\includegraphics[width=0.7\linewidth]{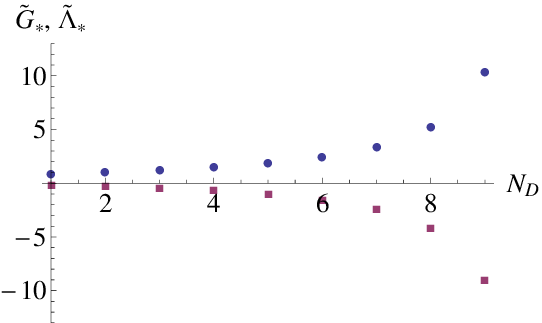}
\caption{\label{GLambdaNF} The values of $\tilde{G}_{\ast}$ (dots) and $\tilde{\Lambda}_{\ast}$ 
(squares) as functions of the number of Dirac fields,
at one loop with type II cutoff.}
\end{figure}

As already observed in \cite{Percacci:2002ie}, 
the effect of fermions is to push $\tilde{G}_{\ast}$ to larger values 
and $\tilde{\Lambda}_{\ast}$ to more negative values, cf. fig.~\ref{GLambdaNF}, see app.~\ref{fpvalues} for detailed values.

At a critical number of fermions $N_D\approx 10.1$,
$\tilde G_*$ goes to $+\infty$ and $\tilde\Lambda_*$ goes to $-\infty$.
This is similar to the behavior seen in the perturbative analysis.
Accordingly fermions have a destabilizing effect on asymptotic safety in gravity, 
reminiscent of a similar effect of fermions on asymptotic freedom in gauge theories.

\begin{figure}[!h]
\includegraphics[width=0.7\linewidth]{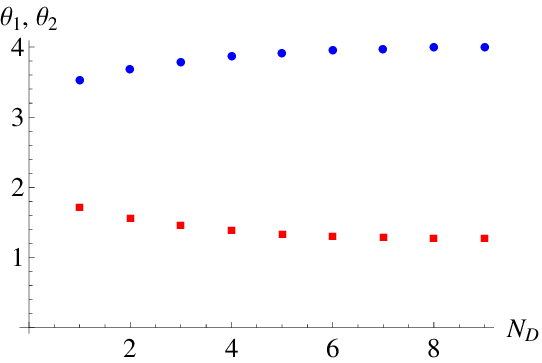}
\caption{\label{thetaND} The critical exponents $\theta_{1,2}$ as functions 
of the number of Dirac fields.}
\end{figure}

\begin{widetext}

\begin{figure}[!h]
{\resizebox{1\columnwidth}{!}
{\includegraphics{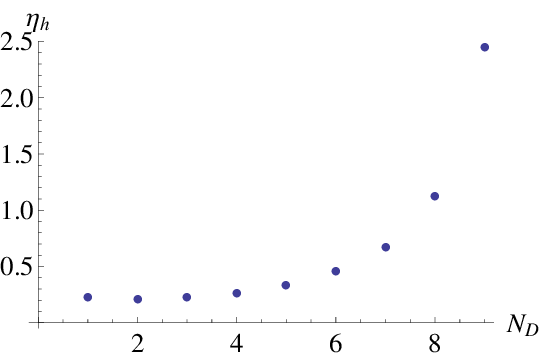}
\includegraphics{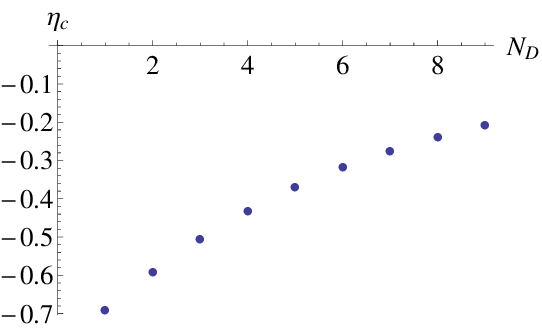}
\includegraphics{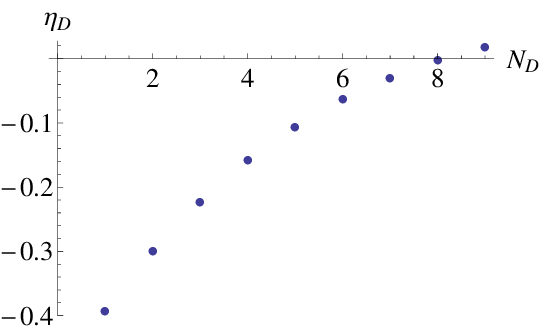}
}
\caption{\label{etaND} The graviton (left), ghost (middle) and fermion (right) anomalous dimensions as functions of the number of fermion fields, all at one loop and with type II cutoff.}
}
\end{figure}

\end{widetext}

Below the critical number the solution to the microscopic \red{equations of motion} at the fixed point -- within our truncation -- is the Euclidean version of anti deSitter space. Thus, AdS/CFT-type dualities might be of use to understand the microscopic gravitational action.

Fermionic fluctuations have only a small effect on the values of the critical exponents, cf.~fig.~\ref{thetaND}. This suggests that fermionic matter does not change the number of relevant directions of background operators. 
In contrast, the graviton anomalous dimension grows, cf.~fig.~\ref{etaND}.
These results show only a very weak scheme-dependence.
The main difference in the RG improved case lies in the fact that the fermionic anomalous dimension remains negative up to the critical value of $N_D$.

\begin{figure}[!h]
\includegraphics[width=0.85\linewidth]{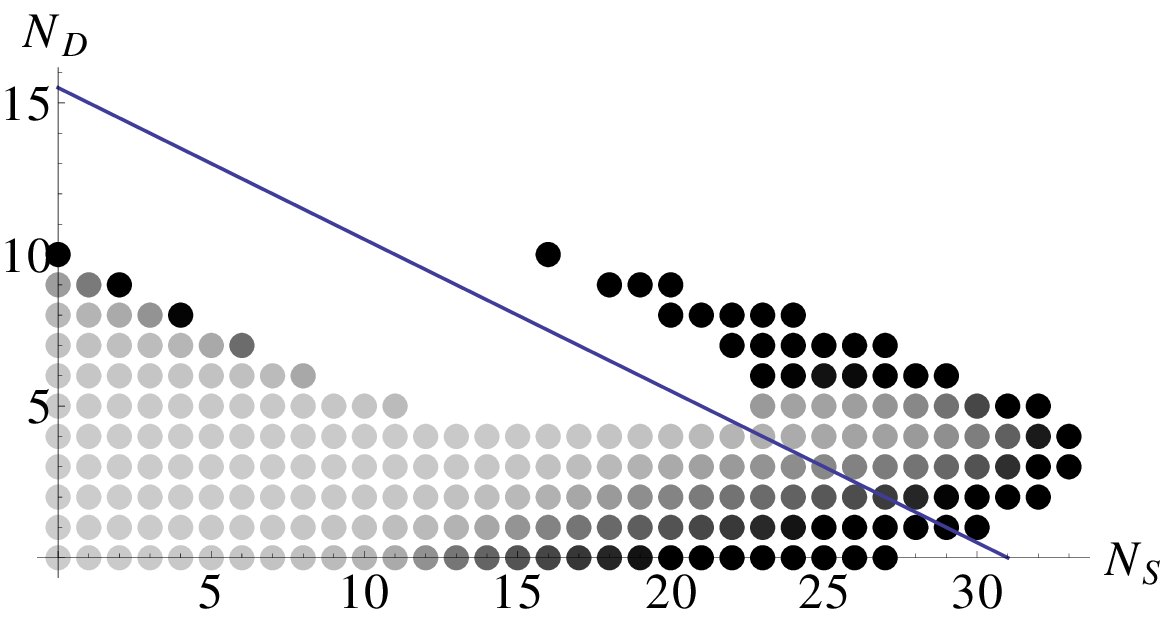}
\caption{\label{exclusionplot} The points in the $N_S$-$N_D$ plane compatible with a gravitational fixed point with two relevant directions for $N_V=0$. The line represents the perturbative bound (\ref{lsing}). Lighter shades of gray mean smaller $\eta_h$; black means $\eta_h>10$.
}
\end{figure}

The main result of our analysis up to this point lies in the existence of a maximum number of fermions and scalars compatible with the gravitational fixed point within our truncation. 
This is true also for combinations of scalars and fermions, as seen in fig.~\ref{exclusionplot}, which shows the existence region of the fixed point in the $N_S$-$N_D$-plane for $N_V=0$.
Note the qualitative agreement with the analysis of the perturbative approximation
in section IV.A. 
We conclude that the inclusion of dynamical matter can fundamentally change a quantum theory of gravity, or even make it inconsistent. It is thus crucial to include realistic matter degrees of freedom in the investigation of the asymptotic-safety scenario for quantum gravity.

\subsubsection{Vector fields}

In contrast to scalars and fermions, we find no bound on the number of vector fields
compatible with a viable gravitational fixed point.
The effect of vector degrees of freedom is always to decrease $\tilde{G}_{\ast}$
and to increase $\tilde{\Lambda}_{\ast}$.
The position of the fixed point and the values of the critical exponents and
anomalous dimensions are shown in figs.~\ref{GLambdaNV},\ref{etaNV},
for $0\leq N_V\leq50$, covering all phenomenologically interesting models.
The behavior is very smooth.
From the point of view of the $N_V$-dependence, however, this is still
a transient range.
For very large $N_V$ all quantities reach the following asymptotic values:

\begin{table}[!h]
\caption{}
\begin{ruledtabular}
\label{mattermodeltab} 
\begin{tabular}{cccccccc}
& $\tilde G_*$ & $\tilde\Lambda_*$ & $\theta_1$ & $\theta_2$ & $\eta_h$ & $\eta_c$ & $\eta_V$
\\\hline
$\lim_{N_V\to\infty}$ & $0$ & $3/8$ & $4$ & $2$ & $9/10$ & $0$ & $0$ 
\end{tabular}
\end{ruledtabular}
\end{table}

This picture holds with small quantitative changes also when the RG improvement 
is taken into account, and with type Ia cutoff.
The most significant difference lies in the fact that the vector anomalous dimension does not change sign even for large $N_V$, when the RG improvement is taken into account.
We therefore believe that the existence of the fixed point
is a true feature of gravity coupled to vector fields.
It will be interesting to see whether the gauge coupling remains asymptotically free
when Yang-Mills is coupled to gravity.

As a preliminary step, we consider the effect of gravitational fluctuations on the beta-function of the gauge coupling in the abelian case. 
In the context of asymptotic safety, this has been considered previously
in \cite{Daum:2009dn,Folkerts:2011jz,Harst:2011zx}. In $d=4$ the beta function is given by
\be
\beta_g = \frac{1}{2}\eta_V g \approx \frac{g^3 \, N_D}{12 \pi^2} - \frac{3}{8 \pi} g\, \tilde{G} + \frac{3}{2 \pi} g \, \tilde{G} \tilde{\Lambda}^2.
\ee
For a small value of the cosmological constant, we observe that gravitational fluctuations lead to an asymptotically free fixed point. Whether this behavior carries over to the QED coupling when defined as in \cite{Toms:2011zza}, is an open question.
As in \cite{Harst:2011zx}, there is also a non-Gau\ss{}ian fixed point, at which the QED coupling is irrelevant, and its value in the infrared can be predicted. A main difference to \cite{Harst:2011zx} lies in the $\tilde{\Lambda}$-dependence of our result, which is quadratic, instead of linear in $\tilde{\Lambda}$.
For $|\tilde{\Lambda}| > \sqrt{0.4}$, the Gau\ss{}ian and the non-Gau\ss{}ian fixed point merge, and only a UV-repulsive Gau\ss{}ian fixed point remains. 
For the matter content of the Standard Model, only the UV-repulsive noninteracting fixed point remains.

\begin{widetext}

\begin{figure}[!h]
{\resizebox{1\columnwidth}{!}
{\includegraphics{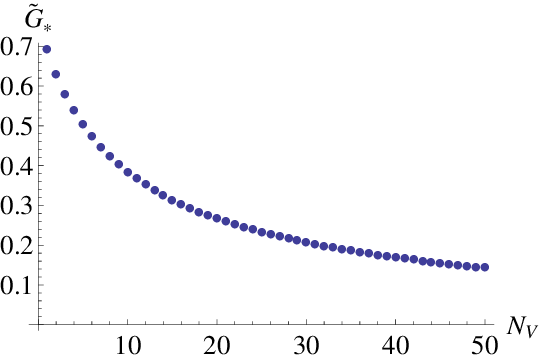}
\includegraphics{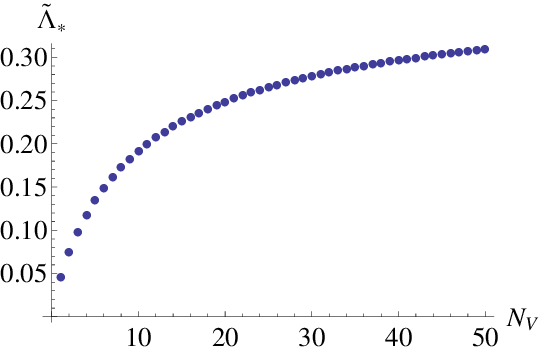}
\includegraphics{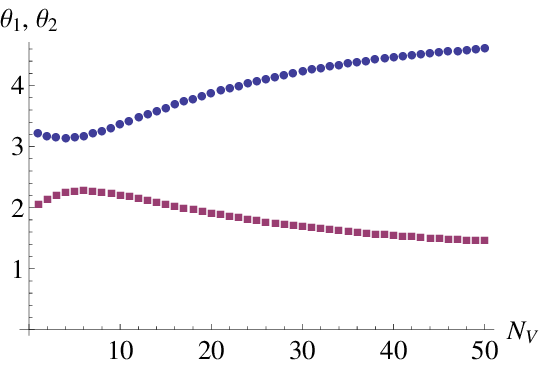}}
\caption{\label{GLambdaNV} Position of the fixed point (left and middle)
and critical exponents (right) as a function of number of vector fields.}
}
\end{figure}

\begin{figure}[!h]
{\resizebox{1\columnwidth}{!}
{\includegraphics{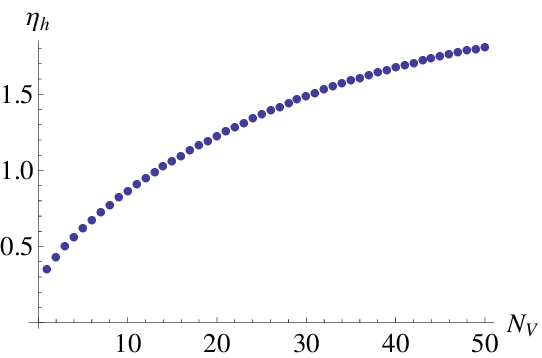}
\includegraphics{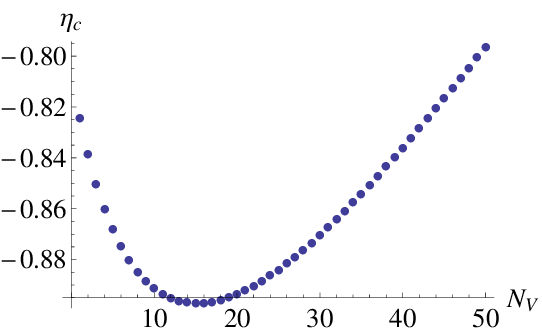}
\includegraphics{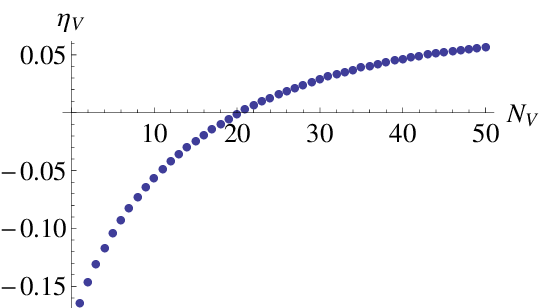}
}
\caption{\label{etaNV} The graviton (left), ghost (middle) and vector (right) anomalous dimensions as functions of the number of vector fields.}
}
\end{figure}

\end{widetext}

\subsubsection{Specific matter models}

A viable gravitational fixed point exists for a small number of matter fields. Increasing the number of matter fields, two mechanisms can remove the fixed point: A first possibility is for $\tilde{G}_{\ast}$ and/or $\tilde{\Lambda}_{\ast}$ to diverge. 
A second mechanism is the collision of fixed points: 
The beta functions admit several zeros, which move in the $\tilde{G}$-$\tilde{\Lambda}$ plane
in dependence of the number of matter fields. 
These fixed points can collide, at which point they move off the real axis to complex values. 
These mechanisms are responsible for the existence of boundaries in the $(N_S, N_D, N_V)$ space.

Before discussing the shape of the boundaries, we will investigate the compatibility 
of specific matter models with the asymptotic safety scenario. 
Note that our results rely on a particular truncation, thus extended truncations 
could result in quantitative changes.
Recall also that we neglect all matter-self-interactions which are present 
in specific matter models. 
In particular we do not distinguish between abelian and non-abelian gauge bosons.

We begin with the standard model (in its original form excluding right-handed neutrinos).
For reasons that have been discussed earlier, we take as our benchmark the one-loop
results obtained with type II cutoff.
These are reported in the first column of the following table.

\begin{table}[!h]
\caption{Standard model matter}
\begin{ruledtabular}\label{mattermodeltab} 
\begin{tabular}{ccccc}
                     & 1L-II    & full-II  & 1L-Ia    &full-Ia  \\\hline
\ $\tilde\Lambda_*$  & $-2.399$ & $-2.348$ & $-3.591$ & $-3.504$ \\\hline
\ $\tilde G_*$       &  $1.762$ &  $1.735$ &  $2.627$ &  $2.580$ \\\hline
\ $\theta_1$         &  $3.961$ & $3.922$  &  $3.964$ &  $3.919$ \\\hline
\ $\theta_2$         &  $1.644$ & $1.651$  &  $2.178$ &  $2.187$ \\\hline
\ $\eta_h$           &  $2.983$ &  $2.914$ &  $4.434$ &  $4.319$ \\\hline
\ $\eta_c$           & $-0.139$ & $-0.129$ & $-0.137$ & $-0.125$ \\\hline
\ $\eta_S$           & $-0.076$ & $-0.072$ & $-0.076$ & $-0.073$ \\\hline
\ $\eta_D$           & $-0.015$ &  $0.004$ & $-0.004$ &  $0.016$ \\\hline
\ $\eta_V$           & $-0.133$ & $-0.145$ & $-0.144$ & $-0.158$ \\\hline
\end{tabular}
\end{ruledtabular}
\end{table}

So the first and most important observation is that the matter content of the
standard model is compatible with the existence of a fixed point.
By first adding, one at the time, the vector fields, then the scalars,
then the fermions, one can convince oneself that this fixed point is
a continuous deformation of the one discussed in section IV.C.1.
The second column shows the properties of the fixed point of the RG improved
beta functions. They are not very different from the one loop results,
as expected from the fact that the anomalous dimensions are not very large.
The other two columns show the properties of the same fixed point when one uses
a type Ia cutoff.
As observed earlier, with this cutoff the allowed region is smaller,
so the standard model is closer to the boundary and this explains why
the couplings are larger.
The variations are typical for the scheme dependence in this approach.
All the evidence leads us to believe that this fixed point is a genuine
feature of the theory and not an artifact of the truncation.

\begin{table}[!h]
\caption{Fixed-point values, critical exponents and anomalous graviton dimension for specific matter content.}
\begin{ruledtabular}\label{mattermodeltab} 
\begin{tabular}{ccccccccc}
{\rm model} & $N_S$ &$N_D$ &$N_V$ & $\tilde{G}_{\ast}$ & $\tilde{\Lambda}_{\ast}$& $\theta_1$& $\theta_2$&$\eta_h$\\
\hline\\
{\rm no matter}     & 0 &  0   &  0& 0.77 &  0.01 & 3.30 & 1.95 & 0.27\\\hline
{\rm SM}            & 4 & 45/2 & 12& 1.76 & -2.40 & 3.96 & 1.64 & 2.98\\\hline
{\rm SM +dm scalar} & 5 & 45/2 & 12& 1.87 & -2.50 & 3.96 & 1.63 & 3.15\\\hline
{\rm SM+ 3 $\nu$'s} & 4 & 24   & 12& 2.15 & -3.20 & 3.97 & 1.65 & 3.71\\\hline
{\rm SM+3$\nu$'s}&{}&{}&{}&{}&{}&{}&{}&{}\\
\rm{ + axion+dm}    & 6 & 24   & 12& 2.50 & -3.62 & 3.96 & 1.63& 4.28\\\hline
{\rm MSSM}&49&61/2&12& -&-&-&-&-\\\hline
{SU(5) GUT}&124&24&24&-&-&-&-&-\\\hline
{\rm SO(10) GUT}&97&24&45&-&-&-&-&-\\\hline
\end{tabular}
\end{ruledtabular}
\end{table}

\begin{figure}[!h]
\includegraphics[width=1\linewidth]{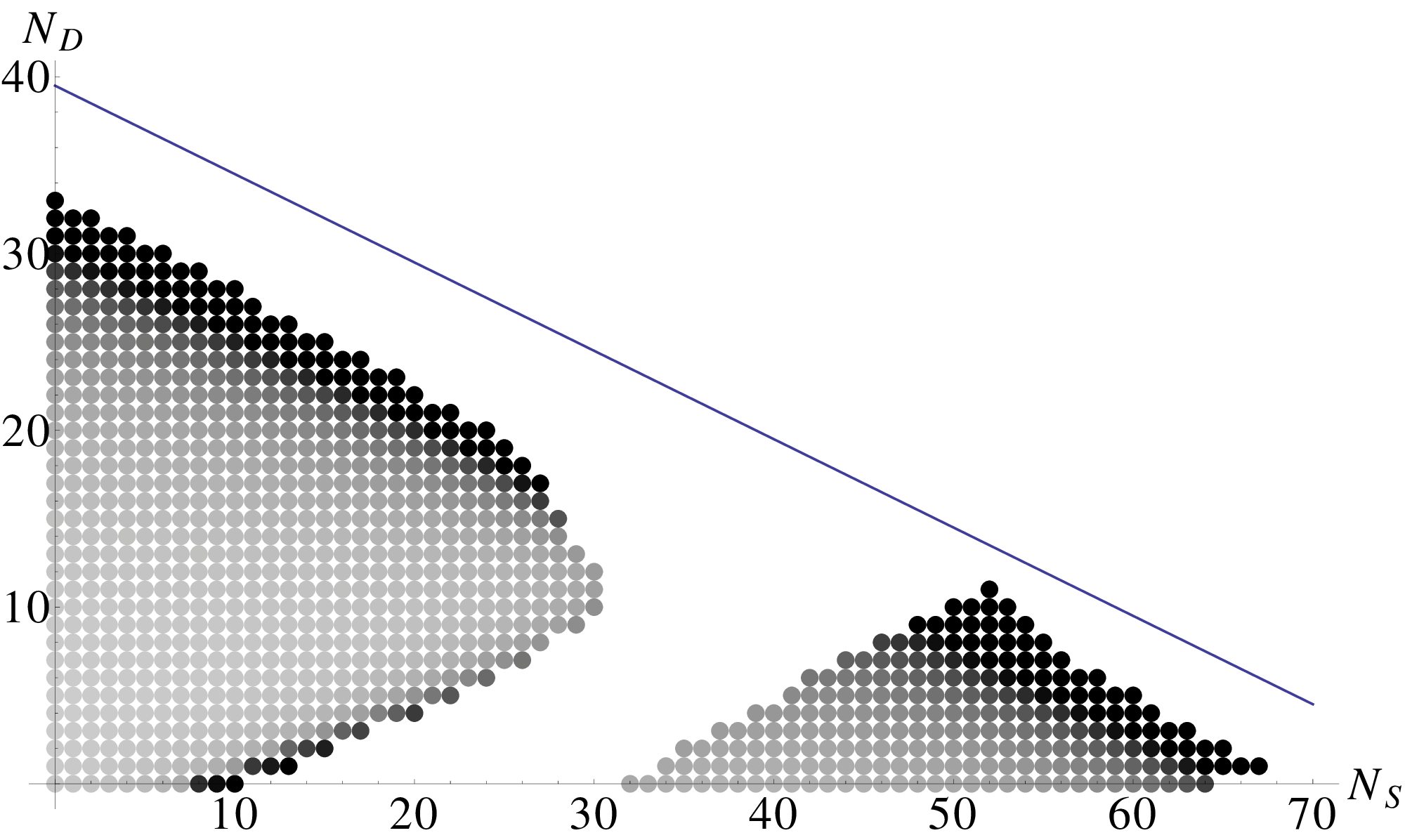}
\caption{\label{eta12plane}The region compatible with the existence of a gravitational fixed point 
with $\tilde G_*>0$ and two attractive directions for $d=4$, $N_V=12$. 
The line represents the perturbative bound (\ref{lsing}). Lighter shades of gray mean smaller $\eta_h$; black means $\eta_h>10$.}
\end{figure}

Theories that go beyond the standard model contain more fields.
So let us consider these models, starting from the ones with fewer fields.
A very minimal extension is a single further scalar field, 
which can be viewed as a model of dark matter \cite{Cline:2013gha,Silveira:1985rk,McDonald:1993ex,Burgess:2000yq}.
This has a small effect, as seen in the third row of table \ref{mattermodeltab}.
In the fourth row we consider a model with three right-handed neutrinos, 
to account for neutrino masses. This has a somewhat larger effect but is
still clearly compatible with asymptotic safety.
In the fifth row we consider a model with three right-handed neutrinos and two scalars,
one of which can be thought of as the axion \cite{Peccei:1977hh, Weinberg:1977ma, Wilczek:1977pj,Essig:2013lka}, the other as dark matter.
This model is still in the allowed region with the type II cutoff,
but if one were to use the more stringent type Ia cutoff it would be quite close to the boundary.
This model is therefore nearly as extended as one can get without adding further gauge fields.
The extent of the allowed region with $N_V=12$ is shown in figure \ref{eta12plane}.

One important example of a model that is beyond the boundary is the MSSM.
In the one loop approximation with type II cutoff there is actually no real solution
with the matter content of the MSSM.
In the RG improved equations, two real solutions with positive $\tilde G_*$ exist,
but they have one positive and one negative critical exponent and are therefore not
a continuous deformation of the no-matter fixed point.
A similar situation holds with type Ia cutoff.
Since they do not appear already in the one loop approximation they are likely 
to be truncation artifacts.
We conclude that the matter content of the MSSM, again neglecting the non-abelian nature of the gauge bosons, is incompatible with the existence of a fixed point.

In the case of GUTs the fermion content is the same as the SM (typically with 
right-handed neutrinos included) and there are more gauge fields,
so one may hope that they are compatible with a fixed point.
In this case, however, it is the large number of scalars that poses a severe challenge.
As examples we consider an $SU(5)$ and an $SO(10)$ model, both with
minimal scalar sectors.
In both cases the fermionic sector consists of three generations including right-handed
neutrinos, {\it i.e.}, a total of 48 Weyl spinors, which count like 24 Dirac spinors.
In the case of SU(5), we consider three scalar multiplets:
one in the adjoint (24 real fields), one in the fundamental
(5 complex fields) and one in the complex 45-dimensional representation.
This sums up to NS = 124.
In the case of SO(10) a minimal scalar sector would contain
the adjoint (45 real fields) the fundamental (10 complex fields)
and one (complex) 16 \cite{Bertolini:2009qj} leading to NS = 97.

The $SU(5)$ model actually has one fixed point with large $\tilde G_*=37$,
large critical exponents $-80$ and $38$ with opposite signs, 
implying that it is not connected to the no-matter fixed point. 
Furthermore  it has a huge anomalous dimension
$\eta_h=84$ which makes it rather unreliable.
The RG improved beta functions again have a single real fixed point
with positive $\tilde G_*=0.21$, but in a very different position and with
very different critical exponents $-5.3$ and $1.4$ which make it impossible
to identify it with the one of the one loop approximation.
This strengthens the suspicion that they are both truncation artifacts.
So while one cannot strictly exclude the existence of a fixed point for this
particular matter content, it is beyond the boundary of our allowed region.
In the case of the $SO(10)$ model this conclusion is even stronger,
since there is no real nontrivial fixed point.
Considering that realistic GUT models have many more scalars than the minimal
models considered here, one can conclude with good confidence that
GUTs with fundamental scalars are incompatible with a gravitational fixed point.

Technicolor-like models \cite{Sannino:2009za}, which dispense with fundamental scalars, 
and instead introduce further fermions and gauge bosons, could very well be compatible 
with a fixed-point scenario for gravity, as larger numbers of vectors also imply 
a larger number of fermions compatible with the fixed point. 

Fig. ~\ref{eta12plane} shows the region in the $N_S$-$N_D$-plane
where a fixed point exists with $\tilde G_*>0$,
$\theta_1,\theta_2>0$ for $N_V=12$, at one loop and with type II cutoff.
In comparison to the perturbative results, the inclusion of the anomalous dimensions 
leads to a more complicated shape of the boundary,
but it remains true that continuous deformations of the fixed point
without matter are only possible in a bounded domain of the plane.
When one increases the number of scalars or fermions at fixed $N_V$
one encounters a singularity, or the fixed point becomes complex.

\begin{widetext}

\begin{figure}[!h]
{\resizebox{0.7\columnwidth}{!}
{
\includegraphics{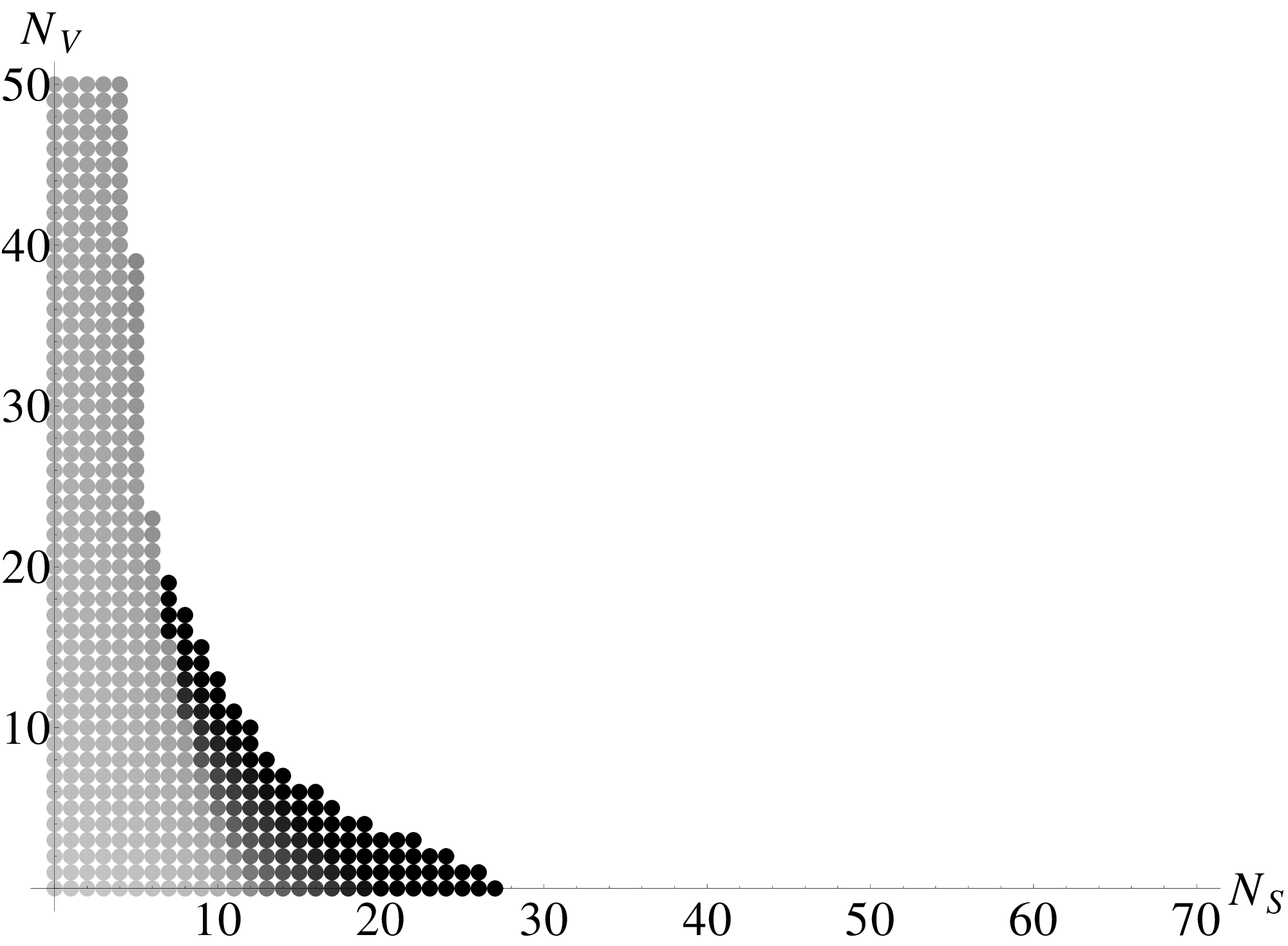}\ \
\includegraphics{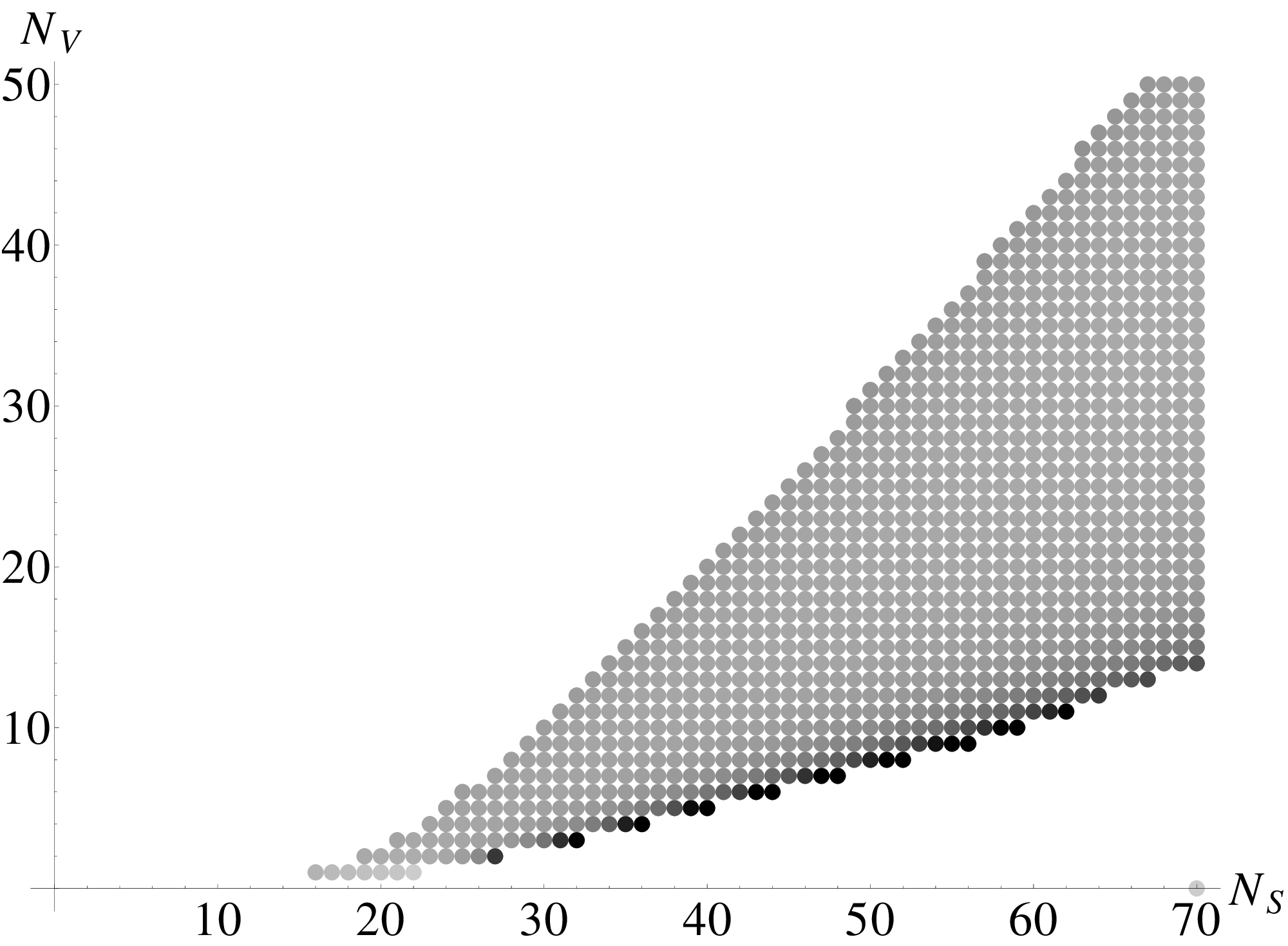}}
\caption{\label{NSNDNV} Left: existence region of the fixed point connected
to the one without matter, in the plane $N_D=0$. \red{Right:} existence region of
a second real fixed point with two relevant directions, in the same plane.
The line $N_D=0$ in fig.~\red{\ref{eta12plane}} corresponds to the line $N_V=12$
in these two figures.}
}
\end{figure}

\end{widetext}

The fixed points in the disconnected island on the right 
cannot be continuously deformed into the one without matter.
Instead, they are the continuation of a fixed point that is complex
in the permitted region connected to the origin, and becomes a pair
of real fixed points for larger number of scalars.
For small $N_V$ the gap closes and there are combinations of matter fields
such that the two fixed points are both real.
This can be \red{seen} in fig. \red{\ref{NSNDNV},} which shows the exclusion plot in the plane $N_D=0$.
(No such phenomena occur in the $N_S=0$ plane.)
Since the second fixed point coexists with the one that we regard as physically significant
in some region of the $(N_S,N_D,N_V)$ space,
it is probably an artifact of the truncation.
Consequently, also in the rest of the space, its significance is doubtful.
More detailed investigations will be necessary to clarify this point.

The shades of grey in figures \ref{exclusionplot}, \ref{eta12plane}, \ref{NSNDNV}
are related to the value of the graviton anomalous dimension,
with darker tones indicating a larger anomalous dimension.
We observe that $\eta_h$ becomes very large ($O(10^3)$) at some
points in the horn of fig. \ref{exclusionplot} and near the
boundary, at small $N_S$.
The restriction $\eta_h>-2$ is automatically satisfied everywhere and
does not add significant restrictions\red{;}
however\red{,} the dark dots indicate that the truncation used is
unreliable.
Our graphs should therefore be taken with a grain of salt,
as the shape and position of the boundary could change
in an extended truncation.

The overall conclusion of this brief investigation
is that asymptotic safety puts very strong restrictions on the matter content.
It is thus interesting to observe that limited, observationally well-motivated extensions of the standard model are compatible with a fixed point for gravity, while models that demand a larger number of degrees of freedom for internal consistency reasons (such as supersymmetric models) are incompatible with the fixed-point scenario. 
The observation of many more fundamental particles at LHC or future accelerators
could therefore pose a severe challenge to the asymptotic safety scenario.

\subsection{The quantum gravity scale with matter}
Although the asymptotic safety scenario aims at a construction of a continuum quantum gravity model, where no fundamental kinematical length scale exists, a quantum-gravity scale will emerge dynamically. This is very similar to QCD, where quantum fluctuations lead to the dynamical generation of $\Lambda_{\rm QCD}$, which is a physical scale at which the behavior of the theory changes drastically. In quantum gravity the transition scale to the fixed-point regime is the dynamically generated quantum-gravity scale. There, the theory changes from the phase in which the dimensionful Newton coupling is constant, to a scale-free regime in which $G(k^2) \sim \frac{1}{k^2}$, which is conjectured to become visible, e.g., in graviton-mediated scattering cross sections  \cite{Litim:2007iu,Gerwick:2011jw,Dobrich:2012nv}. A priori, this scale could take any value, but has been found to be close to the Planck scale in previous studies of the Einstein-Hilbert truncation \cite{Reuter:2004nx}. Note that this notion of a quantum gravity scale differs from that discussed, e.g., in \cite{Calmet:2008df}, where a quantum gravity scale is defined by $\tilde{G} \sim 1$. These two scales differ. The latter can be understood as a scale where quantum gravity effects in general become important. The former is a scale pertaining to the notion of asymptotic safety, and can be thought of as a scale at which predictions from asymptotic safety will differ from other quantum gravity theories.

Trajectories passing very closely to the Gau\ss{}ian fixed point before approaching the UV fixed point  \cite{Reuter:2004nx} exist also under the inclusion of matter. We thus expect to find a fine-tuned trajectory where the gravitational couplings take on their measured values in the infrared. On trajectories similar to this highly fine-tuned one, all quantities then clearly show the dynamical emergence of a scale at which the fixed-point regime is reached.
We fix the dimensionless Newton coupling and cosmological constant to fixed values $\tilde{G}_0, \tilde{\Lambda}_0$ at a given IR scale, and then numerically integrate the RG flow towards the UV. 
 We observe that scalars seem to have little effect on the transition scale, whereas fermions shift this scale towards larger values. There are two competing effects at work here: If the fixed-point coordinates of the NGFP are further away from the GFP, the flow takes up more \red{``RG-time''} until it reaches the fixed point, if the critical exponents are unchanged. If the critical exponents change also, they also alter the amount of \red{``RG-time''} necessary to reach the fixed point. Since the effect of fermions is to induce a considerable shift in the fixed-point values towards larger $\tilde{G}$ and more negative $\tilde{\Lambda}$, they shift the QG scale towards higher values.
In the case of the Standard Model, the effect is less pronounced than in the case with fermions only. In our evaluation, we found a shift of the transition scale to the fixed-point regime by a factor of approximately 10.

Our study suggests that the dynamically generated quantum gravity scale is not independent of the existing matter degrees of freedom, as has also been observed in \cite{Calmet:2008df}. If a shift to higher scales is confirmed beyond our truncation, discovering phenomenological imprints of asymptotically safe quantum gravity might become even more challenging. It would be interesting to include the effect of the running Newton coupling into numerical calculations in the strong-gravity regime, such as those performed in \cite{East:2012mb}; see also \cite{Basu:2010nf,Falls:2010he} for the discussion of black-hole production in asymptotic safety.

\subsection{Higher-dimensional cases}
Large extra dimensions have a number of theoretical motivations, and have been shown to be compatible with asymptotically safe gravity in the Einstein-Hilbert truncation \cite{Fischer:2006fz}  and under the inclusion of fourth-order derivative operators \cite{Ohta:2013uca}. While extra dimensions are not necessary for the consistency of the model, they seem well compatible with it.
Phenomenological implications have been studied in \cite{Litim:2007iu,Gerwick:2011jw,Dobrich:2012nv}.
Experimentally, the best upper bounds on their radius come from recent LHC results, see, e.g., \cite{ATLAS:2012ky,Aad:2012cy}.

\begin{figure}[!h]
\includegraphics[width=\linewidth]{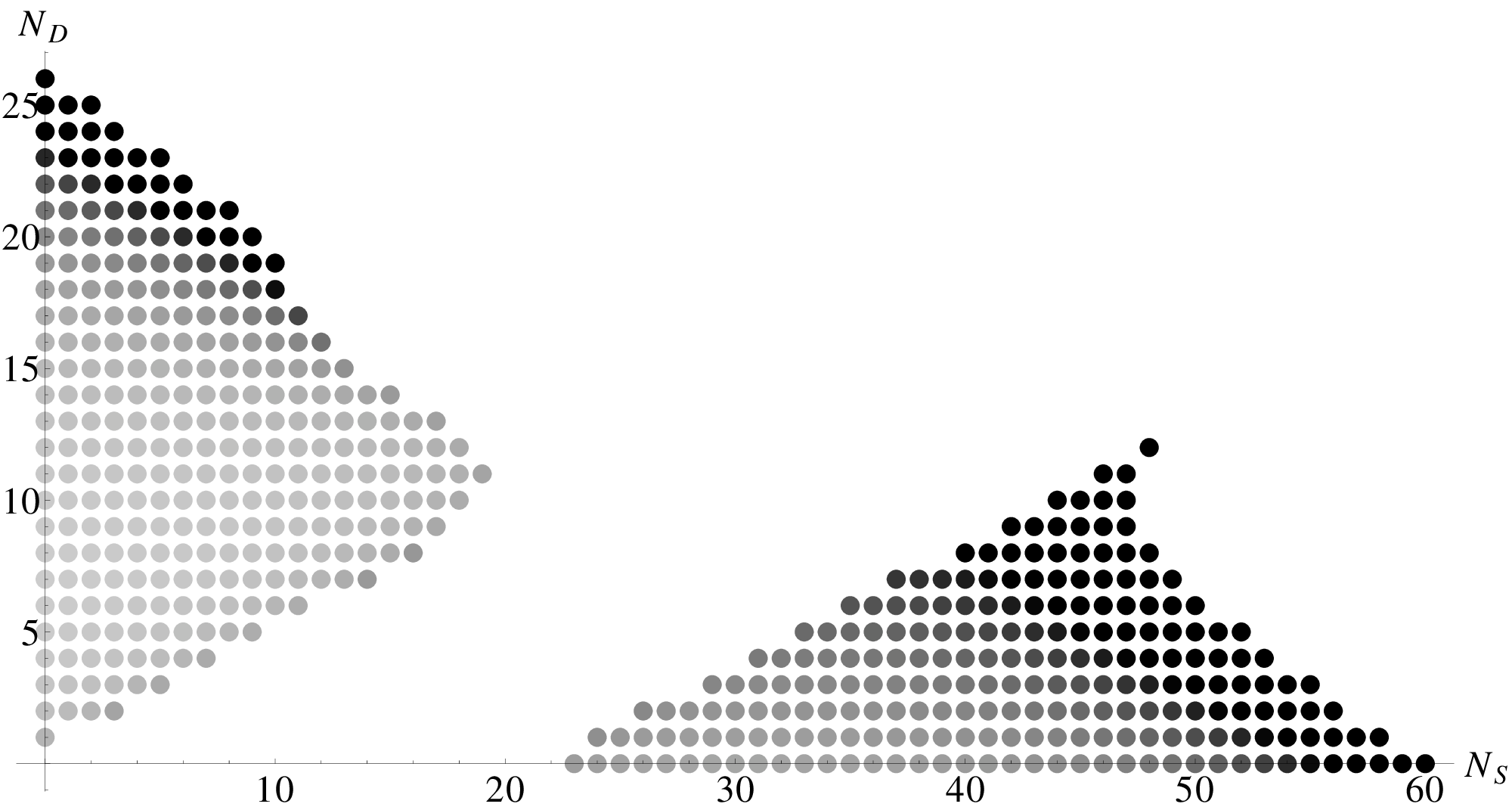}\\
\includegraphics[width=\linewidth]{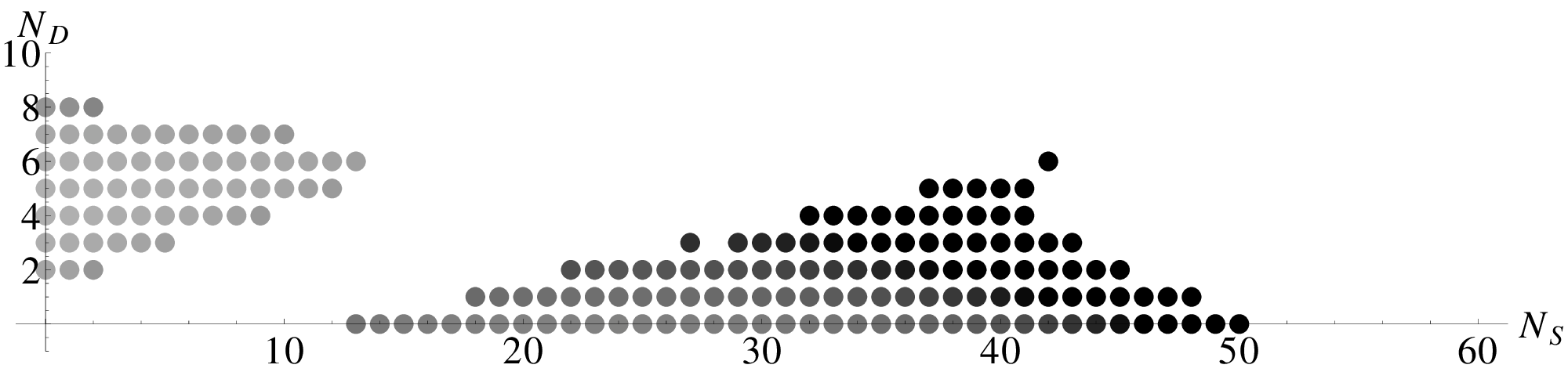}
\caption{\label{5dexcl} The region compatible with the existence of a gravitational fixed point 
with $\tilde G_*>0$ and two attractive directions for $d=5$ (top) and $d=6$ (bottom) and $N_V=12$. 
Lighter shades of gray mean smaller $\eta_h$; black means $\eta_h>10$.
}
\end{figure}

While the extra dimensions have to be compactified in a realistic setting, we can neglect the effect of compactifications here: At momentum scales much higher than the inverse compactification radius, the difference between a continuum of momentum modes and a discrete set of Kaluza-Klein modes has no effect. 

The allowed regions for $d=5,6$ and $N_V=12$, at one loop and with type II cutoff,  are shown in fig.~\ref{5dexcl}.
We see that the standard model would still be (barely) compatible with a fixed point in $d=5$
but it is not in $d=6$.
It would be incompatible also in $d=5$ if we used the type Ia cutoff.
This case is therefore marginal and needs further investigation.

It appears that the existence of sufficiently many matter fields poses a new restriction on the number of dimensions compatible with asymptotic safety. While a pure gravitational fixed point exists in these dimensions, the observed matter degrees of freedom tend to destroy it, 
at least within our truncation. 
Whether this changes if only gravity can propagate into the extra dimensions would have to be re-examined.

\subsection{Anomalous dimension}
In the scale-free fixed-point regime, the value of the anomalous dimension can be related to the momentum-dependence of the propagator as follows
\be
P(p) \sim \frac{1}{\left(p^2\right)^{1- \eta/2}}.
\ee
It is interesting to observe that while a negative value of $\eta$ implies a UV suppression of the propagator, a positive value corresponds to a UV-enhancement. 
For the background-field anomalous dimension, $\eta_N =-2$ arises as a fixed-point requirement, and could be read as a UV suppression. 
In contrast, the fluctuation field anomalous dimension without matter implies a rather weak UV enhancement of the graviton propagator. 
Scalar and fermion fields push $\eta_h$ into the large positive region,
and vectors also tend to increase it, at least as long as $N_V$ is not too large.
The result is that for the Standard Model we have a rather large positive 
anomalous dimension, corresponding to a strong enhancement.

A large anomalous dimension $\eta_h$ implies that for a more detailed understanding of the imprints of asymptotic safety on  scattering cross sections 
in graviton-mediated scattering processes, the inclusion of vertex anomalous dimensions is necessary, as the UV scaling of the propagator alone implies an increase of the scattering cross section at high energies.

Quantitative changes are expected for $\eta_h$, when momentum-dependent higher-order correlation functions for the graviton are included in the truncation, and it remains to be investigated, whether $\eta_h<0$ then, as one might naively expect for the unitarization.
Making the simplifying assumption that graviton-matter vertices with one graviton and two matter fields do not get any explicit renormalization, the quantum-fluctuation-induced scaling of those vertices is determined purely by the anomalous dimensions. For instance, the dimensionless coupling $g_{h\phi\phi}$ of an operator of the form $h_{\mu \nu} \partial_{\mu} \phi \partial_{\nu} \phi$, will then scale as $g_{h \phi \phi} \sim k^{\eta_S + \frac{\eta_h}{2}}$. For graviton-mediated scattering processes, we  encounter a divergence of the scattering cross section with the center-of-mass-energy at tree-level, if quantum-gravity-induced renormalization effects are not taken into account. Here we observe, that, upon an identification of $k^2$ with the center-of-mass energy $s$, we would conclude that a negative anomalous dimension for matter and gravity improves the situation, as there is the additional scaling $s^{2\eta_S+ \eta_h}$. Additionally, the graviton propagator scales nontrivially with an additional $\frac{1}{p^{-\eta_h}}$. The combined effect could even lead to a fall-off of the scattering cross-section, depending on the size of the anomalous dimensions. 
Matter fluctuations tend to drive $\eta_h$ to positive values, which is not the correct sign expected for the unitarization. Note that this expectation is subject to the scale identification 
$k \sim \sqrt{s}$, which might not capture the quantum effects correctly \cite{Anber:2011ut}.

\section{Conclusions}

In this work, we have reexamined the compatibility of minimally coupled
matter degrees of freedom with the asymptotic safety scenario for gravity.
This issue had been addressed before \cite{Percacci:2002ie, Percacci:2003jz},
but advances in our understanding in the intervening ten years \red{have led} to
corrections and improvements.
Our present treatment differs in two crucial ways.
One is the way the cutoff is implemented on fermions:
In \cite{Percacci:2002ie, Percacci:2003jz} 
the cutoff was chosen to be a function of $-\Box$ (so-called type I cutoff),
whereas here the cutoff is chosen to be a function of $-\Box+\frac{R}{4}$.
As discussed in \cite{Dona:2012am}, the correct procedure is to impose a cutoff on the
eigenvalue of the Dirac operator itself, and this is equivalent to the type II procedure.
The second difference is in the treatment of the anomalous dimension of the graviton.
In order to close the flow equations, the approximation $\eta_h=\eta_N\equiv\partial_t G/G$ 
was made earlier.
However, the effective average action $\Gamma_k(g_{\mu\nu},h_{\mu\nu})$
is a functional of two fields: the background metric and the fluctuation,
and the wave function renormalization of the fluctuation has 
a different scale dependence from $1/G$, which appears as a prefactor in the
background part of the action.
We have followed here the calculation of the graviton anomalous dimension of
\cite{Codello:2013fpa}, but adopting a slightly different definition of the anomalous dimension. Additionally, we have explicitly calculated the matter anomalous dimensions, 
which also enter the gravitational beta functions.
As a result of these changes, the allowed region in the $(N_S,N_D,N_V)$ space
is quite different from that of \cite{Percacci:2002ie, Percacci:2003jz}.

Our main finding is that within the Einstein-Hilbert truncation for the gravitational 
degrees of freedom, and with minimally coupled matter, there are upper limits 
on the allowed number of scalar and fermionic degrees of freedom. 
Increasing the number of vector fields leads to slightly weaker bounds. 
Focusing on models of particular interest, we find that the standard model matter content 
is compatible with an appropriate fixed point. 

Small extensions of the SM, e.g., the inclusion of right-handed neutrinos, an axion
and a scalar dark matter particle, are still compatible,
but big enlargements are highly problematic.
In spite of the increase in the number of gauge fields, realistic GUTs have too many scalars.
Things would improve assuming that the gauge symmetry is broken dynamically,
but we are not aware of detailed models of this type.
The MSSM does not show a viable fixed point within our approximations. 
Models with a larger number of vector degrees of freedom, such as technicolor models, can 
accommodate a larger number of fermions and still be compatible with the existence 
of the fixed point.

Going to a larger number of dimensions, we find that the allowed region for the matter content shrinks, and there is no viable gravitational fixed point compatible with the standard model matter content in $d=6$, while the case $d=5$ is in the balance. This indicates that while the gravitational dynamics allows for a fixed point in any number of dimensions \cite{Fischer:2006fz,Ohta:2013uca}, matter dynamics is sensitive to the dimensionality, 
and tend to destroy the gravitational fixed point above $d=4$. 
Extrapolating the trend from d=4, 5, 6, we do not expect that the standard model is compatible with a nontrivial gravitational fixed point in any number of extra dimensions.
Our work thus indicates that a realistic model of asymptotically safe gravity, which includes dynamical matter, disfavors scenarios with universal extra dimensions.  
In the future, it could be interesting to examine this in extended truncations, and to study different models for extra dimensions. For instance, our conclusion could change if only gravity can propagate into the extra dimensions, but matter fields cannot.

We also examined the effect of matter degrees of freedom on the quantum gravity scale. In the asymptotic safety scenario this is the dynamically generated transition scale to the fixed-point regime. Physical observables will presumably exhibit a change of behavior at this scale, even though there is not strictly speaking a phase transition. We find that, fixing the IR value of the Newton coupling and the cosmological constant to small positive values and integrating toward the UV, the fixed-point scale is moved to higher values under the inclusion of matter, e.g., in the standard model case. We trace this back to the fact that for the standard model, the fixed-point coordinates are further away from the Gau\ss{}ian fixed point, in the vicinity of which the RG flow towards the UV starts at low scales. Having a higher transition scale implies that 
effects of asymptotic safety might become more challenging to detect\red{.} 

The limitations of our work should be evident.
First of all, we have restricted ourselves to the cosmological and Newton coupling.
This is partly justified by the fact that in extended truncations
they are confirmed to be the most relevant ones.
Perhaps some further support for this approximation comes from
one loop calculations, where the (universal) four-derivative terms do not yield 
new constraints on matter, and where there exists at least one cutoff scheme
where the same is true of all higher derivative terms.

Our calculations involve rather large uncertainties, so most results
should be taken as broad trends rather than precise statements.
The beta functions are obtained by off-shell calculations and are gauge-dependent.
We have used throughout the Feynman-de Donder gauge $\alpha=1$.
Regarding the ``scheme''-dependence, 
most results have been given at one loop and with a type II cutoff on all degrees of freedom, 
but we have thoroughly studied also the case with a type Ia cutoff on the gravitons,
and (with both cutoffs) the RG-improved flow equations.
In the perturbative approximation we have found that using the type Ia cutoff
leads to a restriction of the allowed region by 12 Dirac fields or 24 scalar fields.
This can be taken as a typical theoretical uncertainty in this type of calculation.
Having given the results for the less restrictive type II cutoff, it is likely
that we have erred by allowing models that are forbidden, rather than the converse.
Detailed calculations with different schemes and/or different gauges will be
necessary to sharpen the boundary of the allowed region.

Another strong limitation is the truncation on the gravitational action.
Even within the context of terms with two derivatives only, due to the natural bi-metric dependence
of the effective average action, there is a difference between the cosmological and Newton
couplings that multiply the background field terms, and the coefficients of the
terms involving powers of the fluctuation $h_{\mu\nu}$.
In most of the literature, the coefficients of the fluctuation terms have been
treated as in the expansion of the Einstein-Hilbert action, thus identifying
them with the cosmological and Newton coupling.
Here we have made a first distinction between the background Newton coupling and
the coefficient of the $p^2 h^2$ term, which we called $Z_h$.
Preliminary results indicate that the next step, where we distinguish between
the background cosmological constant and a ``mass'' term for the graviton,
do not lead to a qualitatively very different picture.
We will return to this extension elsewhere.
Still further extensions would correspond to reading off the strength of the gravitational fluctuation coupling from the graviton three- or four- point functions.

Perhaps most importantly, we have neglected all matter self-interactions.
From \cite{Eichhorn:2012va} \red{it} is known that quantum gravity fluctuations 
induce momentum-dependent matter self-interactions, which couple back into 
the anomalous dimensions. We expect that our boundaries in the $(N_S, N_D, N_V)$-space 
will change under a corresponding extension of the truncation, 
which is however much beyond the scope of the present work.

In spite of these limitations, our work clearly shows that ``matter matters'' 
in asymptotically safe quantum gravity. 
Asymptotic safety might not be compatible with arbitrary extensions of the Standard Model; 
e.g., supersymmetric extensions and higher dimensions seem to be disfavoured. 
This opens a new route to obtain experimental guidance in the construction 
of a viable model of quantum gravity: 
The discovery of many new fundamental matter fields at the LHC or future colliders
could potentially lead to a situation that is theoretically inconsistent with asymptotic safety.
\\

{\emph{Acknowledgements:}
We thank A. Codello, G. D'Odorico and C. Pagani for sharing some of their results prior to publication.
P.D. thanks the Perimeter Institute for hospitality during this work.
This research was supported in part by Perimeter Institute for Theoretical Physics. Research at Perimeter Institute is supported by the Government of Canada through Industry Canada and by the Province of Ontario through the Ministry of Research and Innovation.}

\begin{appendix}
\section{Projection rules}\label{projectionrules}
From the truncation \Eqref{gravity_action}, \Eqref{abelian_action} we deduce the following projection rules onto the anomalous dimensions. Here and in the following \red{we} employ the following conventions for the Fourier-transform:
\bea
h_{\mu \nu}(x) &=& \int_p \tilde{h}_{\mu \nu}(p)e^{i p \cdot x},\qquad
A_{\mu}(x)= \int_p \tilde{A}_{\mu}(p)e^{i p \cdot x},\nonumber\\
\phi(x)&=& \int_p \tilde{\phi}(p)e^{i p \cdot x}, \nonumber\\
\psi(x)&=& \int_p \tilde{\psi}(p)e^{i p \cdot x}, \qquad
\bar{\psi}(x)= \int_p \tilde{\bar{\psi}}(p) e^{-i p \cdot x}.
\eea
In the following, we will drop the tilde to denote the Fourier transform. Whether we refer to the field in position space or in momentum space will become clear from the argument of the field.
The projection rules then read
\bea
\partial_t Z_h&=& 32 \,\pi\, G Z_N \frac{4}{d(3d-2)} \frac{1}{2d}\frac{\partial}{\partial q_{\mu}} \frac{\partial}{\partial q_{\mu}} \nonumber\\
&{}& K_{\alpha\beta\rho\sigma}
\int_p \frac{\delta}{\delta h_{\alpha \beta}(-q)} \frac{\delta}{\delta h_{\rho \sigma}(p)} \partial_t \Gamma_k, \label{etahproj}\\
\partial_t Z_V &=&\frac{1}{2d^2}\frac{\partial}{\partial q_{\mu}} \frac{\partial}{\partial q_{\mu}}   \left(\delta_{\kappa \lambda} + (\alpha -1) \frac{q_{\kappa} q_{\lambda}}{q^2} \right)\cdot \nonumber\\
&{}& \cdot\int_{p}\frac{\delta}{\delta A_{\kappa}(q)}\frac{\delta}{\delta A_{\lambda}(-p)} \partial_t \Gamma_k, \label{etaAproj}
\\
\partial_t Z_S&=& \frac{1}{2d}\frac{\partial}{\partial q_{\mu}} \frac{\partial}{\partial q_{\mu}} \int_p \frac{\delta}{\delta \phi(p)} \frac{\delta}{\delta \phi(q)} \partial_t \Gamma_k,\\
\partial_t Z_D&=& -\frac{1}{2d} \frac{1}{2^{d/2}} \frac{\delta^{ij}}{N_D} {\rm tr_D}\frac{\partial}{\partial q_{\mu}} \frac{\partial}{\partial q_{\mu}}  \slashed{q}\cdot \nonumber\\
&{}& \cdot\int_p \frac{\overset{\rightarrow}{\delta}}{\delta \bar{\psi}^i(q)} \partial_t \Gamma_k \frac{\overset{\leftarrow}{\delta}}{\delta \psi^j(q)}.
\eea

where $\rm tr_D$ denotes a trace over Dirac indices.
\section{Vertices}

Here we list the matter-graviton vertices arising from the kinetic terms. 
We then use the following notation:
\be
\Gamma_{k\, \Phi_a \Phi_b}^{2}(p,q)= \frac{\overset{\rightarrow}{\delta}}{\delta \Phi^T_a(-p)} \Gamma_k \frac{\overset{\leftarrow}{\delta}}{\delta \Phi_b(q)}.
\ee
Herein it only plays a role for the Grassmann valued fermion fields whether the derivative acts from the left or from the right. $\Phi_a(-p)$ denotes a superfield $\Phi^T_a(-p) = (h_{\mu \nu}(-p), A_{\mu}^T(-p), \phi(-p), \psi^T(-p), \bar{\psi}(p) )$ and $\Phi_b(p) = (h_{\mu \nu}(p), A_{\mu}(p), \phi(p), \psi(p), \bar{\psi}^T(-p) )$.

\begin{widetext}
We denote the antisymmetrisation $V_{[\mu \nu]} = \frac{1}{2}(V_{\mu \nu}- V_{\nu \mu})$ and symmetrisation $V_{(\mu \nu)}=\frac{1}{2}(V_{\mu \nu}+ V_{\nu \mu})$.
We obtain:
\bea
\Gamma_{k\, A_{\alpha}h_{\kappa \lambda}}(p,q)&=&\frac{Z_V}{2} \sqrt{32 \pi G}A_{\beta}(p-q) \Bigl(p\cdot(q-p)\delta_{\alpha \beta}\delta_{\kappa \lambda} - p_{\beta} (q_{\alpha}- p_{\alpha})\delta_{\kappa \lambda} - p_{\kappa}(q_{\lambda}- p_{\lambda})\delta_{\alpha \beta} - p_{\lambda}(q_{\kappa}- p_{\kappa})\delta_{\alpha \beta} \nonumber\\
&{}&+p_{\kappa}(q_{\alpha}-p_{\alpha})\delta_{\lambda \beta} + p_{\lambda}(q_{\alpha}- p_{\alpha})\delta_{\beta \kappa }- p\cdot(q-p)\delta_{\alpha \kappa} \delta_{\beta \lambda} - p\cdot (q-p) \delta_{\alpha \lambda} \delta_{\beta \kappa} + p_{\beta}(q_{\kappa}-p_{\kappa}) \nonumber\\
&{}&+p_{\beta} (q_{\lambda}-p_{\lambda}) \delta_{\alpha \kappa} \Bigr),\\
\Gamma_{k\, h_{\kappa \lambda} A_{\alpha}}(p,q)&=& \Gamma_{k\, A_{\alpha}h_{\kappa \lambda}} (-q,-p),\\
\Gamma_{k\, h_{\kappa \lambda} \phi}(p,q)&=& - \frac{Z_S}{2}\sqrt{32 \pi G} \left(\delta_{\kappa \lambda} (p \cdot q -q^2) - p_{\kappa} q_{\lambda} - p_{\lambda}q_{\kappa} + 2 q_{\kappa}q_{\lambda}\right) \phi(p-q),\\
\Gamma_{k\, \phi h_{\kappa \lambda}}(p,q)&=&\Gamma_{k\, h_{\kappa \lambda} \phi}(-q,-p),\\
\Gamma_{k\, h_{\kappa \lambda} \psi}(p,q)&=&\frac{Z_D}{4}\sqrt{32 \pi G} \bar{\psi}(p-q)\left( \delta_{\kappa \lambda} \gamma^\rho(2 q_\rho - p_\rho) - \gamma_\lambda (2q_\kappa +p_\kappa)\right),\\
\Gamma_{k\, h_{\kappa \lambda} \bar{\psi}}(p,q)&=&\frac{Z_D}{4}\sqrt{32 \pi G} \left( \delta_{\kappa \lambda} \gamma^\rho(2 q_\rho - p_\rho) - \gamma_\lambda (2q_\kappa +p_\kappa)\right)\psi (p-q).
\eea
\bea
\Gamma_{k\, h_{\alpha \beta}h_{\gamma \delta}}\vert_{A_{\mu} = \psi = \bar{\psi}= h= 0}(p,q)&=& - \frac{Z_S}{16} 32 \pi G \int_l \phi(p-q-l) \phi(l) \Bigl(2 \left(\delta_{\alpha \beta} \delta_{\gamma \delta}- (\delta_{\alpha \gamma} \delta_{\beta \delta} + \delta_{\alpha \delta} \delta_{\beta \gamma}) \right)(p-q-l) \cdot l \nonumber \\
&{}& -2 \delta_{\alpha \beta} 2 (p_{(\gamma}- q_{(\gamma} - l_{(\gamma}) l_{\delta)} - 2 \delta_{\gamma \delta} 2 (p_{(\alpha}- q_{(\alpha} -l_{(\alpha})l_{\beta)}\nonumber\\
&{}& +2 \left(2 \delta_{\gamma (\alpha} (p_{\beta)}- q_{\beta)} - l_{\beta)})l_{\delta} + 2 \delta_{\delta (\alpha} (p_{\beta) }-q_{\beta)}-l_{\beta)})l_{\gamma}\right)\Bigr)
\eea

For the matter contributions to the graviton anomalous dimension, we also require the second functional derivative of $\Gamma_k$ with respect to the matter fields, and expanded to first order in the metric fluctuation field, where $h(p) = \delta^{\mu \nu}h_{\mu \nu}(p)$.
\bea
\Gamma_{k\, A_{\alpha}A_{\beta}} (p,q)&=& \frac{Z_V}{4} \sqrt{32 \pi G}\left( \frac{1}{2}h(p-q)  \delta^{\mu \nu} - 2 h^{\mu \nu}(p-q)\right) \Bigl(p_{\mu}q_{\nu}\delta_{\alpha \beta} - 2 p_{\mu}q_{\alpha}\delta_{\nu \beta} + p \cdot q \delta_{\alpha \mu} \delta_{\beta \nu} + q_{\mu}p_{\nu} \delta_{\alpha \beta} \nonumber\\
&{}&- 2 q_{\mu} p_{\beta} \delta_{\alpha \nu} + p \cdot q \delta_{\alpha \nu} \delta_{\beta \mu} \Bigr),\\
\Gamma_{k\, \phi \phi}(p,q)&=& \frac{Z_S}{2} \sqrt{32 \pi G\, }h_{\mu \nu}(p-q) \left(\delta_{\mu \nu} p \cdot q - p_{\mu}q_{\nu}-p_{\nu}q_{\mu}\right),
\eea
\bea
\Gamma_{k\, \bar{\psi} \psi}(p,q)&=& \frac{Z_D}{4} \sqrt{32 \pi G\, }
\left( h(p+q) \gamma^\rho \left(p_\rho + q_\rho \right) - h^{\mu\nu}(p+q)\gamma_\nu \left(p_\mu + q_\nu \right)\right).
\eea

For the two tadpole diagrams with internal (external) photons and external (internal) gravitons, we require the following:
\bea
&{}&\frac{\delta}{\delta A_{\chi}(p_1)} \frac{\delta}{\delta A_{\xi}(-p_2)} \frac{\delta}{\delta h_{\tau \eta}(-p)} \frac{\delta}{\delta h_{\gamma \delta}(q)}\Gamma_k\\
 &=&- \frac{Z_V}{8} 32 \pi G \Bigl(\Bigr[\frac{1}{2} \left(\delta_{\alpha \tau} \delta_{\beta \eta} + \delta_{\alpha \eta} \delta_{\beta \tau} \right)\frac{1}{2} \left(\delta_{\gamma \mu}\delta_{\delta \nu}+ \delta_{\gamma \nu}\delta_{\delta \mu} \right)
+\frac{1}{2} \left(\delta_{\alpha \gamma} \delta_{\beta \delta}
+ \delta_{\alpha \delta} \delta_{\beta \gamma}\right) \frac{1}{2} \left(\delta_{\tau \mu} \delta_{\eta \nu}
+ \delta_{\tau \nu} \delta_{\eta \mu}\right)\Bigl] \cdot \nonumber
\\
&{}& \Bigl\{2 p_{1[\kappa} \delta_{\lambda]\chi} 2(p_{[\rho}- q_{[\rho}- p_{1[\rho})\delta_{\sigma]\xi}  +2(p_{[\kappa} -q_{[\kappa}- p_{1[\kappa})\delta_{\lambda]\xi} 2p_{1[\rho}\delta_{\sigma]\chi}\Bigr\} \cdot \nonumber\\
&{}& \Bigl[ \left( -\frac{1}{2} \delta_{\alpha \mu} \delta_{\beta \nu} + \frac{1}{4} \delta_{\alpha \beta} \delta_{\mu \nu}\right) \delta_{\kappa \rho} \delta_{\lambda \sigma} - 2 \delta_{\kappa \rho} \delta_{\alpha \beta} \delta_{\lambda \mu} \delta_{\sigma \nu} + 4 \delta_{\beta \nu} \delta_{\alpha \kappa} \delta_{\mu \rho} \delta_{\lambda \sigma} +2 \delta_{\mu \kappa} \delta_{\nu \rho} \delta_{\alpha \lambda} \delta_{\beta \sigma}\Bigr]\Bigr) \delta^{d}(p-q-p_1-p_2).
\nonumber
\\
&{}& \frac{\delta}{\delta \phi(p_1)}\frac{\delta}{\delta\phi(-p_2)} \frac{\delta}{\delta h_{\mu \nu}(-p)} \frac{\delta}{\delta h_{\kappa \lambda}(q)}\Gamma_k
\\
&=&- \frac{Z_S}{16} 32 \pi G\Bigl(-4 \delta_{\mu \nu} \delta_{\kappa \lambda}p_1 \cdot p_2+4 (\delta_{\mu \kappa} \delta_{\nu \lambda}+ \delta_{\mu \lambda} \delta_{\nu \kappa}) p_1 \cdot p_2  +4 \delta_{\mu \nu} 2 p_{1(\kappa}p_{2\lambda)} \nonumber\\
&{}&-8\left(\delta_{\nu (\kappa} p_{1 \lambda) }p_{2\mu}+ \delta_{\mu (\kappa} p_{1\lambda)}p_{2\nu} + \delta_{\kappa (\mu} p_{1 \nu)}p_{2 \lambda} + \delta_{\lambda (\nu}p_{1\mu)} p_{2\kappa}\right)\Bigr)\delta^d(-p_1+p_2-p+q).
\eea
Herein the derivatives with respect to the external fields are part of the projection rule \Eqref{etahproj} and \Eqref{etaAproj}. 
From the fact that these vertices are at most first order in the graviton momenta, it is clear that there is no contribution to $\eta_h$ from the \red{tadpole} diagrams.

For the fermionic part, we will show the expansion to second order in the gravitons here, from which the expressions for the vertices can be derived straightforwardly:
\bea
\Gamma_{k}\Big|_{\phi=0=A}&=&-32 \pi G \frac{Z_D}{2}\int_{p,q,l} \bar{\psi}(p+q+l) \Bigl( \left( -\frac{1}{2}h_{\kappa \lambda}(p)h_{\kappa \lambda}(q) + \frac{1}{4}h(p)h(q)\right) (\slashed{p}+ \slashed{q}) + \gamma_{\kappa} h_{\kappa \lambda}(p) h_{\lambda \tau}(q) \left(\frac{3}{4}l_{\tau} + \frac{1}{2}p_{\tau} + \frac{1}{4}q_{\tau} \right) \nonumber\\
&{}& \gamma_{\kappa}h_{\kappa \tau}(q) h(p) \left(- \frac{1}{2} l_{\tau} - \frac{1}{4}p_{\tau} - \frac{1}{4}q_{\tau} \right)+ \frac{1}{4}h(p)\slashed{q} h(q) - \frac{3}{16} \left( \slashed{p}+ \slashed{q}\right) h_{\kappa \lambda}(p) h_{\kappa \lambda}(q) \nonumber\\
&{}&- \frac{1}{8} \gamma^{\tau} \gamma^{\kappa} \gamma^{\lambda} h_{\mu \kappa}(p) h_{\mu \lambda}(q)q_{\tau}\Bigr) \psi(l)
\eea

We obtain the following (inverse) propagators, and also give their inverse for the case of nontrivial tensor structures:
\bea
\Gamma_{kA_{\mu}A_{\nu}}^{(2)}\Big|_{h=0}&=&Z_V \left(p^2 \delta_{\mu \nu} - \frac{\alpha -1}{\alpha} p_{\mu}p_{\nu} \right) \left(1+r_k(p^2) \right)\delta^d(p-q),\\
\left(\Gamma_{kA_{\mu}A_{\nu}}^{(2)}\Big|_{h=0}\right)^{-1}&=& \frac{1}{Z_V p^2 (1+r_k(p^2))} \left(\delta_{\mu \nu} + (\alpha -1) \frac{p_{\mu}p_{\nu}}{p^2} \right),\\
\Gamma_{kh_{\mu\nu}h_{\kappa \lambda}}^{(2)}\Big|_{A=h=\phi= \psi=0}&=& Z_h (p^2 - 2\lambda+R_k(p^2)) \Bigl(\frac{1}{2} \delta_{\mu \kappa} \delta_{\nu \lambda} + \delta_{\mu \lambda} \delta_{\nu \kappa} - \frac{1}{2} \delta_{\mu \nu} \delta_{\kappa \lambda}\Bigr)\delta^d(p-q),\\
\left(\Gamma_{kh_{\mu\nu}h_{\kappa \lambda}}^{(2)}\Big|_{A=h=\phi= \psi=0}\right)^{-1}&=&\frac{1}{Z_h(p^2-2\lambda + R_k(p^2) )}\Bigl(\frac{1}{2}\delta_{\mu \kappa} \delta_{\nu \lambda} + \delta_{\mu \lambda} \delta_{\nu \kappa} - \frac{1}{d-2} \delta_{\mu \nu} \delta_{\kappa \lambda}\Bigr),\\
\Gamma_{k \phi \phi }^{(2)}\Big|_{h=0} &=& Z_S \left( p^2+R_k(p^2) \right) \delta^d(p-q)\\
\Gamma_{k \bar{\psi} \psi}^{(2)}\Big|_{h=0}&=& Z_D \gamma^\mu p_\mu \left(1+r_k(p^2) \right)\delta^d(p-q)\\
\eea
%


\section{Fixed-point values}\label{fpvalues}
\begin{table*}[!htbp]
\caption{Selected fixed-point values as a function of the number of matter fields for $N_V=0$
for type II cutoff and one-loop approximation}
\begin{ruledtabular}\label{fptab}
\begin{tabular}{cccccccccccc}
 $N_S$ &$N_D$  & $\tilde{G}_{\ast}$ & $\tilde{\Lambda}_{\ast}$& $\eta_h$& $\eta_c$&$\eta_S$&$\eta_D$&$\eta_{V}$&$\theta_1$&$\theta_2$\\ \hline
 0 & 0 & 0.7725 & 0.01046 & 0.2690 & -0.8065 & -0.3384 & -0.5014 & -0.1843 & 3.299 & 1.951 \\
 0 & 1 & 0.8969 & -0.08969 & 0.2297 & -0.6898 & -0.2982 & -0.3920 & -0.2092 & 3.531 & 1.725 \\
 0 & 2 & 1.056 & -0.2205 & 0.2185 & -0.5898 & -0.2634 & -0.2995 & -0.2284 & 3.688 & 1.573 \\
 0 & 3 & 1.262 & -0.3954 & 0.2309 & -0.5040 & -0.2331 & -0.2220 & -0.2426 & 3.796 & 1.468 \\
 0 & 4 & 1.538 & -0.6361 & 0.2681 & -0.4309 & -0.2068 & -0.1578 & -0.2520 & 3.871 & 1.395 \\
 0 & 6 & 2.491 & -1.505 & 0.4598 & -0.3169 & -0.1647 & -0.06275 & -0.2596 & 3.958 & 1.315 \\
 0 & 8 & 5.239 & -4.111 & 1.133 & -0.2367 & -0.1336 & -0.00196 & -0.2565 & 3.999 & 1.289 \\
 0 & 10 & 118.9 & -114.0 & 30.91 & -0.1806 & -0.1107 & 0.0356 & -0.2474 & 4.018 & 1.293\\ \hline
 1 & 0 & 0.7702 & 0.03151 & 0.2938 & -0.8652 & -0.3606 & -0.5485 & -0.1830 & 3.239 & 2.008 \\
 1 & 1 & 0.8993 & -0.06720 & 0.2429 & -0.7368 & -0.3165 & -0.4271 & -0.2117 & 3.501 & 1.746 \\
 1 & 2 & 1.068 & -0.1981 & 0.2251 & -0.6268 & -0.2785 & -0.3244 & -0.2343 & 3.674 & 1.572 \\
 1 & 3 & 1.291 & -0.3765 & 0.2341 & -0.5324 & -0.2454 & -0.2380 & -0.2513 & 3.792 & 1.451 \\
 1 & 4 & 1.596 & -0.6282 & 0.2711 & -0.4519 & -0.2167 & -0.1664 & -0.2630 & 3.872 & 1.368 \\
 1 & 6 & 2.718 & -1.593 & 0.4831 & -0.3268 & -0.1707 & -0.06092 & -0.2730 & 3.963 & 1.276 \\
 1 & 8 & 6.595 & -5.062 & 1.384 & -0.2398 & -0.1371 & 0.005561 & -0.2704 & 4.004 & 1.245 \\
 \hline
  2 & 0 & 0.7666 & 0.05282 & 0.3247 & -0.9312 & -0.3853 & -0.6022 & -0.1805 & 3.173 & 2.077 \\
 2 & 1 & 0.8999 & -0.04399 & 0.2592 & -0.7894 & -0.3368 & -0.4671 & -0.2134 & 3.467 & 1.774 \\
 2 & 2 & 1.077 & -0.1744 & 0.2332 & -0.6681 & -0.2952 & -0.3528 & -0.2400 & 3.658 & 1.574 \\
 2 & 3 & 1.319 & -0.3559 & 0.2380 & -0.5640 & -0.2589 & -0.2563 & -0.2604 & 3.788 & 1.435 \\
 2 & 4 & 1.659 & -0.6192 & 0.2743 & -0.4751 & -0.2275 & -0.1761 & -0.2748 & 3.874 & 1.338 \\
 2 & 6 & 3.000 & -1.704 & 0.5122 & -0.3371 & -0.1772 & -0.05816 & -0.2881 & 3.970 & 1.231 \\
 2 & 8 & 9.058 & -6.797 & 1.841 & -0.2422 & -0.1407 & 0.01501 & -0.2862 & 4.010 & 1.195 \\
\hline
%
 
  4 & 0 & 0.7563 & 0.09571 & 0.4122 & -1.090 & -0.4440 & -0.7339 & -0.1704 & 3.026 & 2.254 \\
 4 & 1 & 0.8950 & 0.004107 & 0.3048 & -0.9145 & -0.3846 & -0.5653 & -0.2136 & 3.392 & 1.853 \\
 4 & 2 & 1.089 & -0.1231 & 0.2557 & -0.7666 & -0.3343 & -0.4230 & -0.2499 & 3.625 & 1.590 \\
 4 & 3 & 1.370 & -0.3089 & 0.2483 & -0.6390 & -0.2907 & -0.3017 & -0.2793 & 3.780 & 1.404 \\
 4 & 4 & 1.794 & -0.5971 & 0.2817 & -0.5292 & -0.2526 & -0.1995 & -0.3015 & 3.882 & 1.273 \\
 4 & 6 & 3.855 & -2.054 & 0.6015 & -0.3582 & -0.1915 & -0.04825 & -0.3251 & 3.985 & 1.120 \\
 4 & 8 & 51.43 & -36.92 & 9.826 & -0.2431 & -0.1478 & 0.04286 & -0.3259 & 4.020 & 1.065\\ \hline
  8 & 0 & 0.7383 & 0.1794 & 0.8066 & -1.588 & -0.6255 & -1.161 & -0.1210 & 3.378 & 2.184 \\
 8 & 1 & 0.8655 & 0.1022 & 0.4964 & -1.282 & -0.5210 & -0.8688 & -0.1930 & 3.304 & 2.084 \\
 8 & 2 & 1.075 & -0.008829 & 0.3483 & -1.053 & -0.4446 & -0.6432 & -0.2566 & 3.565 & 1.701 \\
 8 & 3 & 1.429 & -0.1880 & 0.2887 & -0.8583 & -0.3804 & -0.4480 & -0.3156 & 3.780 & 1.374 \\
 8 & 4 & 2.101 & -0.5257 & 0.3025 & -0.6842 & -0.3232 & -0.2723 & -0.3698 & 3.919 & 1.110 \\
 8 & 6 & 15.61 & -7.352 & 1.928 & -0.3848 & -0.2244 & 0.02737 & -0.4595 & 4.023 & 0.6906\\ \hline
 12 & 0 & 1.079 & 0.2548 & 3.054 & -3.733 & -1.419 & -2.948 & 0.02049 & 8.304 & 1.143 \\
 12 & 1 & 0.8633 & 0.1927 & 1.092 & -2.000 & -0.7832 & -1.482 & -0.1250 & 4.317 & 1.803 \\
 12 & 2 & 1.025 & 0.1061 & 0.6074 & -1.544 & -0.6265 & -1.050 & -0.2269 & 3.769 & 1.868 \\
 12 & 3 & 1.385 & -0.03850 & 0.4000 & -1.235 & -0.5262 & -0.7345 & -0.3289 & 3.872 & 1.458 \\
 12 & 4 & 2.340 & -0.3808 & 0.3418 & -0.9582 & -0.4419 & -0.4270 & -0.4542 & 4.034 & 0.9425\\ \hline
  16 & 0 & 3.859 & 0.2206 & 6.645 & -10.58 & -4.089 & -8.063 & -0.3468 & 30.15 & 17.53 \\
 16 & 1 & 1.752 & 0.2355 & 3.737 & -5.292 & -2.031 & -4.096 & -0.08671 & 12.68 & 3.306 \\
 16 & 2 & 1.105 & 0.1987 & 1.499 & -2.651 & -1.035 & -1.976 & -0.1490 & 5.901 & 1.587 \\
 16 & 3 & 1.299 & 0.1010 & 0.7452 & -1.914 & -0.7782 & -1.295 & -0.2903 & 4.572 & 1.607 \\
 16 & 4 & 2.183 & -0.1369 & 0.4589 & -1.483 & -0.6491 & -0.8086 & -0.4970 & 4.434 & 0.9745 \\
\hline
 20 & 0 & 10.26 & 0.1761 & 10.17 & -21.67 & -8.547 & -15.79 & -1.724 \
& 122.9 & 47.25 \\
 20 & 1 & 5.036 & 0.1910 & 6.032 & -11.55 & -4.526 & -8.542 & -0.7432 \
& 43.76 & 15.74 \\
 20 & 2 & 2.613 & 0.2030 & 3.669 & -6.428 & -2.505 & -4.812 & -0.3326 \
& 18.32 & 4.817 \\
 20 & 3 & 1.650 & 0.1812 & 1.850 & -3.584 & -1.411 & -2.625 & -0.2666 \
& 8.739 & 1.771 \\
 20 & 4 & 1.968 & 0.06072 & 0.8730 & -2.463 & -1.016 & -1.605 & \
-0.4608 & 6.113 & 1.221 \\
 20 & 8 & 1650. & -12.49 & 176.0 & -23.52 & -13.99 & 2.856 & -29.73 & \
7.2+332.1 i & 7.2-332.1 i \\
\hline 
 24 & 0 & 33.42 & 0.1290 & 19.88 & -55.77 & -22.43 & -38.78 & -7.014 \
& 766.4 & 152.0 \\
 24 & 1 & 13.95 & 0.1338 & 9.037 & -23.82 & -9.564 & -16.65 & -2.886 \
& 173.1 & 41.16 \\
 24 & 2 & 7.377 & 0.1407 & 5.276 & -13.02 & -5.211 & -9.156 & -1.491 \
& 64.35 & 13.79 \\
 24 & 3 & 4.311 & 0.1426 & 3.241 & -7.675 & -3.070 & -5.410 & -0.8655 \
& 28.15 & 4.824 \\
 24 & 4 & 3.219 & 0.1018 & 1.798 & -4.759 & -1.935 & -3.224 & -0.7182 \
& 14.31 & 1.691 \\
 24 & 6 & 219.5 & -2.802 & 9.883 & -14.80 & -8.124 & -1.105 & -14.67 \
& 31.28+97.45 i & 31.28-97.45 i. \\
\hline 
 28 & 1 & 56.02 & 0.06934 & 19.80 & -72.50 & -29.83 & -47.62 & -13.03 \
& 1429. & 166.8 \\
 28 & 2 & 24.94 & 0.05064 & 8.180 & -30.04 & -12.44 & -19.39 & -5.878 \
& 307.9 & 40.31 \\
 28 & 3 & 16.35 & 0.01115 & 4.455 & -17.11 & -7.177 & -10.64 & -3.901 \
& 124.1 & 11.98 \\
 28 & 4 & 21.68 & -0.1880 & 2.599 & -13.03 & -5.773 & -6.798 & -4.789 \
& 84.79 & 4.494 \\
 28 & 6 & 380.9 & -1.301 & 15.43 & -55.95 & -28.68 & -12.76 & -43.49 \
& 1715. & 209.7 

\end{tabular}
\end{ruledtabular}
\end{table*}
%
%
%
\begin{table*}
\caption{Selected fixed-point values as a function of the number of matter fields for $N_V=12$}
\begin{ruledtabular}\label{fptabNV12} 
\begin{tabularx}{\textwidth}{ccccccccccc}
 $N_S$ &$N_D$  & $\tilde{G}_{\ast}$ & $\tilde{\Lambda}_{\ast}$& $\eta_h$& $\eta_c$&$\eta_S$&$\eta_D$&$\eta_{V}$&$\theta_1$&$\theta_2$\\ \hline
0 & 0 & 0.3536 & 0.2078 & 0.9516 & -0.8949 & -0.3480 & -0.6732 & \
-0.04175 & 3.484 & 2.159 \\
 0 & 1 & 0.3697 & 0.1796 & 0.8528 & -0.7960 & -0.3135 & -0.5820 & \
-0.06051 & 3.450 & 2.173 \\
 0 & 2 & 0.3878 & 0.1493 & 0.7944 & -0.7132 & -0.2845 & -0.5060 & \
-0.07581 & 3.483 & 2.126 \\
 0 & 3 & 0.4076 & 0.1163 & 0.7634 & -0.6421 & -0.2596 & -0.4411 & \
-0.08837 & 3.539 & 2.059 \\
 0 & 4 & 0.4290 & 0.08031 & 0.7522 & -0.5801 & -0.2377 & -0.3850 & \
-0.09867 & 3.600 & 1.991 \\
 0 & 6 & 0.4767 & -0.002089 & 0.7710 & -0.4772 & -0.2010 & -0.2933 & \
-0.1138 & 3.705 & 1.878 \\
 0 & 8 & 0.5312 & -0.1010 & 0.8288 & -0.3961 & -0.1718 & -0.2228 & \
-0.1232 & 3.782 & 1.798 \\
 0 & 12 & 0.6664 & -0.3653 & 1.033 & -0.2803 & -0.1289 & -0.1265 & \
-0.1307 & 3.877 & 1.712 \\
 0 & 16 & 0.8564 & -0.7652 & 1.367 & -0.2054 & -0.1002 & -0.06836 & \
-0.1297 & 3.925 & 1.681 \\
 0 & 20 & 1.152 & -1.418 & 1.920 & -0.1555 & -0.08034 & -0.03269 & \
-0.1247 & 3.952 & 1.675 \\
 0 & 24 & 1.693 & -2.647 & 2.961 & -0.1211 & -0.06616 & -0.01033 & \
-0.1182 & 3.968 & 1.681 \\
 0 & 28 & 3.045 & -5.762 & 5.597 & -0.09672 & -0.05572 & 0.003974 & \
-0.1115 & 3.978 & 1.693 \\
 0 & 32 & 13.01 & -28.87 & 25.12 & -0.07889 & -0.04781 & 0.01325 & \
-0.1049 & 3.984 & 1.706\\ \hline 
 1 & 0 & 0.3552 & 0.2173 & 1.021 & -0.9533 & -0.3690 & -0.7241 & \
-0.03473 & 3.577 & 2.097 \\
 1 & 1 & 0.3711 & 0.1893 & 0.8988 & -0.8432 & -0.3307 & -0.6226 & \
-0.05570 & 3.491 & 2.154 \\
 1 & 2 & 0.3892 & 0.1592 & 0.8262 & -0.7526 & -0.2990 & -0.5392 & \
-0.07264 & 3.497 & 2.128 \\
 1 & 3 & 0.4094 & 0.1265 & 0.7863 & -0.6755 & -0.2720 & -0.4687 & \
-0.08652 & 3.542 & 2.068 \\
 1 & 4 & 0.4313 & 0.09075 & 0.7693 & -0.6086 & -0.2484 & -0.4080 & \
-0.09790 & 3.598 & 2.000 \\
 1 & 6 & 0.4804 & 0.008583 & 0.7825 & -0.4983 & -0.2093 & -0.3093 & \
-0.1146 & 3.702 & 1.882 \\
 1 & 8 & 0.5368 & -0.09061 & 0.8391 & -0.4119 & -0.1781 & -0.2338 & \
-0.1251 & 3.781 & 1.797 \\
 1 & 12 & 0.6779 & -0.3578 & 1.048 & -0.2890 & -0.1327 & -0.1311 & \
-0.1337 & 3.876 & 1.706 \\
 1 & 16 & 0.8779 & -0.7671 & 1.397 & -0.2101 & -0.1025 & -0.06983 & \
-0.1328 & 3.925 & 1.672 \\
 1 & 20 & 1.194 & -1.446 & 1.982 & -0.1580 & -0.08181 & -0.03259 & \
-0.1276 & 3.951 & 1.667 \\
 1 & 24 & 1.788 & -2.760 & 3.116 & -0.1224 & -0.06711 & -0.009480 & \
-0.1209 & 3.967 & 1.674 \\
 1 & 28 & 3.369 & -6.309 & 6.173 & -0.09735 & -0.05635 & 0.005141 & \
-0.1138 & 3.977 & 1.686 \\
 1 & 32 & 22.11 & -48.62 & 42.58 & -0.07911 & -0.04824 & 0.01453 & \
-0.1069 & 3.984 & 1.700\\ \hline 
 2 & 0 & 0.3573 & 0.2269 & 1.105 & -1.020 & -0.3929 & -0.7822 & \
-0.02643 & 3.704 & 2.011 \\
 2 & 1 & 0.3726 & 0.1991 & 0.9532 & -0.8956 & -0.3496 & -0.6678 & \
-0.05004 & 3.553 & 2.121 \\
 2 & 2 & 0.3908 & 0.1692 & 0.8633 & -0.7958 & -0.3148 & -0.5758 & \
-0.06887 & 3.522 & 2.122 \\
 2 & 3 & 0.4111 & 0.1368 & 0.8125 & -0.7118 & -0.2854 & -0.4989 & \
-0.08423 & 3.550 & 2.073 \\
 2 & 4 & 0.4335 & 0.1013 & 0.7887 & -0.6396 & -0.2601 & -0.4331 & \
-0.09681 & 3.599 & 2.008 \\
 2 & 6 & 0.4840 & 0.01945 & 0.7950 & -0.5212 & -0.2181 & -0.3267 & \
-0.1154 & 3.700 & 1.886 \\
 2 & 8 & 0.5426 & -0.07991 & 0.8500 & -0.4288 & -0.1849 & -0.2457 & \
-0.1271 & 3.779 & 1.797 \\
 2 & 12 & 0.6898 & -0.3501 & 1.064 & -0.2982 & -0.1367 & -0.1362 & \
-0.1368 & 3.876 & 1.699 \\
 2 & 16 & 0.9007 & -0.7691 & 1.428 & -0.2151 & -0.1050 & -0.07137 & \
-0.1361 & 3.925 & 1.664 \\
 2 & 20 & 1.239 & -1.477 & 2.050 & -0.1606 & -0.08333 & -0.03244 & \
-0.1307 & 3.951 & 1.658 \\
 2 & 24 & 1.894 & -2.887 & 3.290 & -0.1238 & -0.06809 & -0.008551 & \
-0.1236 & 3.967 & 1.665 \\
 2 & 28 & 3.772 & -6.991 & 6.890 & -0.09796 & -0.05699 & 0.006398 & \
-0.1162 & 3.977 & 1.679 \\
 2 & 32 & 74.50 & -162.4 & 143.1 & -0.07931 & -0.04867 & 0.01589 & \
-0.1090 & 3.984 & -1.694\\ \hline 
 3 & 0 & 0.3601 & 0.2367 & 1.210 & -1.097 & -0.4205 & -0.8497 & \
-0.01653 & 3.873 & 1.898 \\
 3 & 1 & 0.3744 & 0.2089 & 1.019 & -0.9543 & -0.3709 & -0.7187 & \
-0.04337 & 3.642 & 2.069 \\
 3 & 2 & 0.3924 & 0.1793 & 0.9068 & -0.8434 & -0.3322 & -0.6164 & \
-0.06439 & 3.564 & 2.107 \\
 3 & 3 & 0.4129 & 0.1472 & 0.8428 & -0.7516 & -0.3001 & -0.5321 & \
-0.08142 & 3.567 & 2.075 \\
 3 & 4 & 0.4357 & 0.1120 & 0.8105 & -0.6733 & -0.2727 & -0.4606 & \
-0.09535 & 3.605 & 2.015 \\
 3 & 6 & 0.4876 & 0.03051 & 0.8085 & -0.5459 & -0.2276 & -0.3457 & \
-0.1159 & 3.700 & 1.891 \\
 3 & 8 & 0.5483 & -0.06895 & 0.8613 & -0.4470 & -0.1921 & -0.2587 & \
-0.1290 & 3.779 & 1.796 \\
 3 & 12 & 0.7021 & -0.3420 & 1.081 & -0.3080 & -0.1410 & -0.1415 & \
-0.1400 & 3.876 & 1.691 \\
 3 & 16 & 0.9248 & -0.7714 & 1.461 & -0.2203 & -0.1075 & -0.07297 & \
-0.1395 & 3.925 & 1.654 \\
 3 & 20 & 1.288 & -1.510 & 2.124 & -0.1633 & -0.08491 & -0.03223 & \
-0.1340 & 3.951 & 1.649 \\
 3 & 24 & 2.014 & -3.032 & 3.487 & -0.1251 & -0.06909 & -0.007529 & \
-0.1265 & 3.966 & 1.657 \\
 3 & 28 & 4.289 & -7.866 & 7.810 & -0.09855 & -0.05765 & 0.007751 & \
-0.1187 & 3.976 & 1.671\\ \hline 
 4 & 0 & 0.3639 & 0.2467 & 1.344 & -1.188 & -0.4533 & -0.9301 & \
-0.004498 & 4.097 & 1.749 \\
 4 & 1 & 0.3767 & 0.2188 & 1.098 & -1.021 & -0.3949 & -0.7766 & \
-0.03548 & 3.764 & 1.995 \\
 4 & 2 & 0.3941 & 0.1894 & 0.9583 & -0.8962 & -0.3514 & -0.6617 & \
-0.0591 & 3.625 & 2.079 \\
 4 & 3 & 0.4147 & 0.1576 & 0.8779 & -0.7952 & -0.3162 & -0.5688 & \
-0.078 & 3.596 & 2.070 \\
 4 & 4 & 0.4378 & 0.1227 & 0.8355 & -0.7101 & -0.2863 & -0.4909 & \
-0.09346 & 3.617 & 2.020 \\
 4 & 6 & 0.4912 & 0.04176 & 0.8234 & -0.5726 & -0.2378 & -0.3665 & \
-0.1163 & 3.702 & 1.895 \\
 4 & 8 & 0.5540 & -0.05771 & 0.8733 & -0.4666 & -0.1999 & -0.2727 & \
-0.1308 & 3.780 & 1.795 \\
 4 & 12 & 0.7149 & -0.3336 & 1.098 & -0.3184 & -0.1456 & -0.1473 & \
-0.1433 & 3.876 & 1.684 \\
 4 & 16 & 0.9504 & -0.7739 & 1.496 & -0.2257 & -0.1102 & -0.07464 & \
-0.1432 & 3.925 & 1.645 \\
 4 & 20 & 1.342 & -1.547 & 2.204 & -0.1661 & -0.08655 & -0.03197 & \
-0.1374 & 3.950 & 1.639 \\
 4 & 24 & 2.152 & -3.198 & 3.713 & -0.1265 & -0.07012 & -0.006403 & \
-0.1296 & 3.966 & 1.648 \\
 4 & 28 & 4.975 & -9.027 & 9.031 & -0.09911 & -0.05831 & 0.009212 & \
-0.1214 & 3.976 & 1.663\\ 
\end{tabularx}
\end{ruledtabular}
\end{table*}
\begin{table*}
\caption{Continue of selected fixed-point values as a function of the number of matter fields for $N_V=12$}
\begin{ruledtabular}
\begin{tabularx}{\textwidth}{ccccccccccc}
 $N_s$ &$N_f$  & $\tilde{G}_{\ast}$ & $\tilde{\Lambda}_{\ast}$& $\eta_h$& $\eta_c$&$\eta_S$&$\eta_D$&$\eta_{V}$&$\theta_1$&$\theta_2$\\ \hline
 6 & 0 & 0.3792 & 0.2682 & 1.777 & -1.452 & -0.5479 & -1.163 & \
0.03061 & 4.831 & 1.285 \\
 6 & 1 & 0.3837 & 0.2391 & 1.321 & -1.188 & -0.4550 & -0.9227 & \
-0.01472 & 4.141 & 1.766 \\
 6 & 2 & 0.3987 & 0.2099 & 1.095 & -1.022 & -0.3970 & -0.7705 & \
-0.04539 & 3.830 & 1.973 \\
 6 & 3 & 0.4186 & 0.1786 & 0.9678 & -0.8967 & -0.3533 & -0.6550 & \
-0.06906 & 3.700 & 2.034 \\
 6 & 4 & 0.4421 & 0.1445 & 0.8976 & -0.7944 & -0.3176 & -0.5610 & \
-0.08811 & 3.669 & 2.016 \\
 6 & 6 & 0.4981 & 0.06476 & 0.8579 & -0.6332 & -0.2609 & -0.4141 & \
-0.1163 & 3.716 & 1.903 \\
 6 & 8 & 0.5655 & -0.03443 & 0.8994 & -0.5107 & -0.2173 & -0.3048 & \
-0.1344 & 3.785 & 1.794 \\
 6 & 12 & 0.7417 & -0.3159 & 1.134 & -0.3415 & -0.1556 & -0.1603 & \
-0.1505 & 3.879 & 1.667 \\
 6 & 16 & 1.006 & -0.7798 & 1.574 & -0.2375 & -0.1160 & -0.07820 & \
-0.1511 & 3.925 & 1.623 \\
 6 & 20 & 1.465 & -1.631 & 2.389 & -0.1719 & -0.09002 & -0.03120 & \
-0.1448 & 3.950 & 1.618 \\
 6 & 24 & 2.499 & -3.615 & 4.282 & -0.1292 & -0.07225 & -0.003794 & \
-0.1362 & 3.964 & 1.629 \\
 6 & 28 & 7.352 & -13.06 & 13.27 & -0.1001 & -0.05966 & 0.01250 & \
-0.1271 & 3.974 & 1.646\\ \hline 
 8 & 0 & 1.439 & 0.2806 & 7.941 & -6.091 & -2.283 & -4.943 & 0.2188 & \
24.48 & 2.515 \\
 8 & 1 & 0.3990 & 0.2606 & 1.712 & -1.441 & -0.5460 & -1.145 & \
0.01753 & 4.831 & 1.366 \\
 8 & 2 & 0.4062 & 0.2307 & 1.303 & -1.188 & -0.4571 & -0.9151 & \
-0.02585 & 4.194 & 1.773 \\
 8 & 3 & 0.4238 & 0.1999 & 1.096 & -1.024 & -0.3994 & -0.7638 & \
-0.05630 & 3.902 & 1.944 \\
 8 & 4 & 0.4469 & 0.1665 & 0.9826 & -0.8971 & -0.3553 & -0.6474 & \
-0.08009 & 3.778 & 1.986 \\
 8 & 6 & 0.5047 & 0.08835 & 0.9014 & -0.7052 & -0.2881 & -0.4716 & \
-0.1149 & 3.750 & 1.907 \\
 8 & 8 & 0.5768 & -0.01008 & 0.9294 & -0.5625 & -0.2376 & -0.3432 & \
-0.1377 & 3.800 & 1.793 \\
 8 & 12 & 0.7705 & -0.2967 & 1.174 & -0.3680 & -0.1670 & -0.1755 & \
-0.1584 & 3.883 & 1.649 \\
 8 & 16 & 1.071 & -0.7871 & 1.662 & -0.2505 & -0.1225 & -0.08206 & \
-0.1600 & 3.926 & 1.599 \\
 8 & 20 & 1.616 & -1.736 & 2.615 & -0.1781 & -0.09378 & -0.03004 & \
-0.1532 & 3.948 & 1.594 \\
 8 & 24 & 2.989 & -4.208 & 5.086 & -0.1318 & -0.07450 & -0.0006048 & \
-0.1435 & 3.962 & 1.607 \\
 8 & 28 & 14.30 & -24.84 & 25.64 & -0.1010 & -0.06105 & 0.01635 & \
-0.1334 & 3.972 & 1.627\\ \hline 
 12 & 1 & 2.290 & 0.2522 & 8.908 & -7.776 & -2.959 & -6.124 & 0.01932 \
& 40.87 &3.654 \\
 12 & 2 & 0.4748 & 0.2771 & 2.506 & -1.951 & -0.7328 & -1.577 & \
0.06171 & 6.469 & 0.7400 \\
 12 & 3 & 0.4482 & 0.2433 & 1.613 & -1.429 & -0.5460 & -1.115 & \
-0.01090 & 4.890 & 1.475 \\
 12 & 4 & 0.4620 & 0.2109 & 1.285 & -1.192 & -0.4627 & -0.8992 & \
-0.05164 & 4.331 & 1.759 \\
 12 & 6 & 0.5186 & 0.1367 & 1.034 & -0.8975 & -0.3599 & -0.6290 & \
-0.1063 & 3.941 & 1.874 \\
 12 & 8 & 0.5994 & 0.04145 & 1.008 & -0.6980 & -0.2899 & -0.4466 & \
-0.1419 & 3.880 & 1.783 \\
 12 & 12 & 0.8349 & -0.2534 & 1.266 & -0.4349 & -0.1956 & -0.2149 & \
-0.1768 & 3.906 & 1.605 \\
 12 & 16 & 1.231 & -0.808 & 1.884 & -0.2816 & -0.1380 & -0.09082 & \
-0.1818 & 3.929 & 1.539 \\
 12 & 20 & 2.053 & -2.047 & 3.272 & -0.1915 & -0.1023 & -0.02591 & \
-0.1736 & 3.944 & 1.535 \\
 12 & 24 & 5.022 & -6.682 & 8.428 & -0.1367 & -0.07935 & 0.008189 & \
-0.1611 & 3.955 & 1.554\\ \hline 
 16 & 3 & 1.142 & 0.2713 & 5.588 & -4.480 & -1.688 & -3.599 & 0.1109 \
& 17.72 & 1.117 \\
 16 & 4 & 0.5235 & 0.2562 & 2.153 & -1.830 & -0.6949 & -1.447 & \
0.01298 & 6.206 & 1.072 \\
 16 & 6 & 0.5397 & 0.1852 & 1.300 & -1.199 & -0.4709 & -0.8813 & \
-0.08425 & 4.515 & 1.702 \\
 16 & 8 & 0.6232 & 0.09557 & 1.133 & -0.8974 & -0.3656 & -0.6043 & \
-0.1405 & 4.116 & 1.739 \\
 16 & 12 & 0.9109 & -0.2027 & 1.379 & -0.5293 & -0.2354 & -0.2729 & \
-0.1994 & 3.964 & 1.546 \\
 16 & 16 & 1.463 & -0.8450 & 2.203 & -0.3215 & -0.1583 & -0.1010 & \
-0.2113 & 3.935 & 1.458 \\
 16 & 20 & 2.884 & -2.651 & 4.523 & -0.2059 & -0.1125 & -0.01750 & \
-0.2012 & 3.933 & 1.455 \\
 16 & 24 & 17.96 & -22.51 & 29.73 & -0.1402 & -0.08463 & 0.02208 & \
-0.1843 & 3.940 & 1.485\\ \hline 
 20 & 4 & 2.630 & 0.2137 & 7.349 & -6.904 & -2.677 & -5.225 & -0.2781 \
& 40.91 & 1.024 \\
 20 & 6 & 0.6084 & 0.2317 & 1.987 & -1.792 & -0.6887 & -1.381 & \
-0.03701 & 6.279 & 1.218 \\
 20 & 8 & 0.6575 & 0.1497 & 1.373 & -1.212 & -0.4833 & -0.8599 & \
-0.1283 & 4.757 & 1.590 \\
 20 & 12 & 1.004 & -0.1441 & 1.528 & -0.6701 & -0.2939 & -0.3632 & \
-0.2277 & 4.110 & 1.459 \\
 20 & 16 & 1.838 & -0.9183 & 2.723 & -0.3750 & -0.1860 & -0.1118 & \
-0.2549 & 3.946 & 1.335 \\
 20 & 20 & 5.200 & -4.374 & 8.019 & -0.2202 & -0.1248 & 0.0003216 & \
-0.2416 & 3.905 & 1.335\\ \hline 
 24 & 6 & 1.775 & 0.2185 & 5.184 & -4.800 & -1.857 & -3.650 & -0.1684 \
& 23.68 & 0.7370 \\
 24 & 8 & 0.7533 & 0.1983 & 1.957 & -1.803 & -0.7042 & -1.343 & \
-0.1021 & 6.619 & 1.216 \\
 24 & 12 & 1.133 & -0.07945 & 1.751 & -0.8962 & -0.3863 & -0.5138 & \
-0.2653 & 4.493 & 1.309 \\
 24 & 16 & 2.644 & -1.114 & 3.838 & -0.4507 & -0.2277 & -0.1170 & \
-0.3305 & 3.953 & 1.116 \\
 24 & 20 & 122.7 & -93.37 & 186.1 & -0.2276 & -0.1394 & 0.04438 & \
-0.3111 & 3.819 & 1.119\\ \hline 
 28 & 12 & 1.392 & -0.02121 & 2.207 & -1.310 & -0.5551 & -0.7911 & \
-0.3317 & 5.638 & 0.967\\ \hline 
 32 & 0 & 1.580 & 1.122 & 2.342 & -0.2386 & 0.05296 & -0.7970 & \
0.8496 & 2.457 & 2.193\\ \hline 
\end{tabularx}
\end{ruledtabular}
\end{table*}

\end{widetext}

\end{appendix}

\FloatBarrier

\end{document}